\documentclass[showpacs,amsmath,reprint,showkeys,superscriptaddress,pra]{revtex4-2}
\usepackage{graphicx,amssymb,amsmath,amsthm,mathrsfs,bbm}
\usepackage{hyperref}
\usepackage{mathtools,xcolor}

\DeclarePairedDelimiter\bra{\langle}{\rvert}
\DeclarePairedDelimiter\ket{\lvert}{\rangle}
\DeclarePairedDelimiterX\braket[2]{\langle}{\rangle}{#1 \delimsize\vert #2}

\newcommand{\Eqref}[1]{(\ref{#1})}
\newcommand{\half}{\frac{1}{2}}
\newcommand{\expo}[1]{\mathrm{e}^{#1}}
\newcommand{\brac}[1]{\left(#1 \right)}

\newcommand{\im}{\mathrm{i}}

\begin{document}
\title{Observing Algebraic Variety of Lee-Yang Zeros in Asymmetrical Systems \\ via a Quantum Probe}


\author{Arijit Chatterjee}
\email{chatterjee.arijit@students.iiserpune.ac.in}
\author{T.~S.~Mahesh}
\email{mahesh.ts@iiserpune.ac.in}
\affiliation{Department of Physics and NMR Research Center, Indian Institute of Science Education and Research, Pune, India}

\author{Mounir Nisse}
\email{mounir.nisse@gmail.com, mounir.nisse@xmu.edu.my}
\author{Yen-Kheng Lim}
\email{yenkheng.lim@gmail.com, yenkheng.lim@xmu.edu.my}
\affiliation{School of Mathematics and Physics, Xiamen University Malaysia, 43900 Sepang, Malaysia}

\begin{abstract}
{Lee-Yang (LY) zeros,  points on the complex plane of physical parameters where the partition function goes to zero, have found diverse applications across multiple disciplines like statistical physics, protein folding, percolation, complex networks etc. However, experimental extraction of the complete set of LY zeros for general asymmetrical classical systems remains a crucial challenge to put those applications into practice. Here, we propose a qubit-based method to simulate an asymmetrical classical Ising system, enabling the exploration of LY zeros at arbitrary values of physical parameters like temperature, internal couplings etc. Without assuming system symmetry, the full set of LY zeros forms an \emph{algebraic variety} in a higher-dimensional complex plane. To determine this \emph{variety}, we project it into sets representing magnitudes (\emph{amoeba}) and phases (\emph{coamoeba}) of LY zeros. Our approach uses a probe qubit to initialize the system and to extract LY zeros without assuming any control over the system qubits. This is particularly important as controlling system qubits can get intractable with the increasing complexity of the system. Initializing the system at an \emph{amoeba} point, \emph{coamoeba} points are sampled by measuring probe qubit dynamics. Iterative sampling yields the entire \emph{algebraic variety}. Experimental demonstration of the protocol is achieved through a three-qubit NMR register. This work expands the horizon of quantum simulation to domains where identifying LY zeros in general classical systems is pivotal. Moreover, by extracting abstract mathematical objects like \emph{amoeba} and \emph{coamoeba} for a given polynomial, our study integrates pure mathematical concepts into the realm of quantum simulations.}
\end{abstract}
	
\maketitle	

\section{Introduction}
\label{sec:Introduction}
It is commonly understood that complex numbers merely play the role of a calculational tool in physics, while real numbers represent any observable quantity. In 1952, Lee and Yang published two seminal papers \cite{Yang:1952be,Lee:1952ig} showing the partition function of a system can become zero at certain points on the complex plane of its physical parameters. These zeros, now known as Lee-Yang (LY) zeros, provide a cohesive understanding of equilibrium phase transition as they correspond to the non-analyticity of free energies. However, it was widely believed that LY zeros can not be observed directly as they occur at complex values of physical parameters, and only when they approach the real axis, their presence gets disclosed as the system goes through a phase transition. Nevertheless, this does not mean that complex LY zeros have nothing to say about the physical system. In fact, determination of LY zeros can play a key role in studying the thermodynamic behavior of complicated many-body systems as they fully characterize the partition function. Apart from that, recent studies \cite{	doi:10.1126/sciadv.abf2447, PhysRevResearch.1.023004} have paved the way for determining universal critical exponents of phase transitions from LY zeros, which is otherwise a computationally difficult problem due to critical slowdown. It was also observed that LY zeros can be employed to understand non-equilibrium phenomena like dynamical phase transitions \cite{PhysRevLett.118.180601},  along with other statistical studies \cite{PhysRevE.97.012115, PhysRevC.72.011901} like percolation   \cite{arndt2001directed} or complex networks \cite{krasnytska2015violation,krasnytska2016partition} and even protein folding \cite{PhysRevLett.110.248101,PhysRevE.88.022710}. Profound links between thermodynamics in the complex plane and dynamical properties of quantum systems have also been discovered \cite{PhysRevLett.110.135704,PhysRevLett.110.050601,PhysRevLett.110.230601,PhysRevLett.110.230602,wei2014phase}  in recent years. All these discoveries have made the determination of LY zeros for  general classical systems a crucial necessity in various disciplines of physics.

In 2012, by representing a classical Ising chain with a small number of spins, Wei and Liu showed \cite{PhysRevLett.109.185701} that the complex LY zeros for such a system can be mapped to the zeros of quantum coherence of an interacting probe.  Using this method, the experimental observation of LY zeros was directly achieved \cite{PhysRevLett.114.010601}. 

This opens up the possibility of using quantum simulation techniques to simulate classical systems and determine their LY zeros with a quantum probe, as demanded in many areas of physics. However, there are two major challenges to  overcome. First of all, the method proposed in \cite{PhysRevLett.109.185701} and successive experiments \cite{PhysRevLett.114.010601,doi:10.1126/sciadv.abf2447} reported observation of LY zeros lying on a single complex plane $\mathbb{C}$. This happens when the partition function is expandable as a univariate polynomial, known as Lee-Yang (LY) polynomial, in terms of the complexified physical parameter. This condition of LY polynomial being univariate is based on a symmetry assumption about the system. For example, in the case of the Ising chain, it assumes that all the spins will experience the same local (complexified) magnetic field. Unfortunately, a general system will not necessarily have this symmetry. Thus, the partition function, in general, is to be expanded as a multivariate LY polynomial in terms of complexified parameters. In the worst case, an $N$ spin Ising chain will have an $N$ variate LY polynomial if all the spins experience different local fields. In such a scenario LY zeros of the system will form an \emph{algebraic variety} \cite{kollar2001simplest}  $V\subset(\mathbb{C}^*)^N$, where $\mathbb{C}^*=\mathbb{C}-\{0\}$.
Therefore, finding a method to experimentally determine the full \emph{algebraic variety} containing LY zeros of a general asymmetrical system is crucial. The second challenge to overcome is a quantum simulator of the classical system should have ways to simulate the system at a wide range of its physical parameters like temperature, internal couplings etc. This flexibility in initialization is crucial to uncover the full set of LY zeros corresponding to all physical situations.  However, the systems under study can get complicated and thus achieving full quantum control over system qubits can get challenging. Therefore, the desired method should not assume much experimental control over the system qubits so that it can be extended to complex systems in the future. 

In this work, we show the direct experimental determination of the \emph{algebraic variety} containing roots of a general multivariate LY polynomial for  asymmetrical Ising type systems. Mathematical developments in last few decades unveil that one can project the \emph{algebraic variety} to sets of coordinate-wise absolute values and arguments, called the \emph{amoeba} \cite{GKZ-94,Viro:2002} and \emph{coamoeba} \cite{Feng:2005gw,Nisse:2013a,Nisse:2013b,Nisse:2017}, respectively. We use qubits to simulate the classical asymmetrical Ising system which can be controlled through another qubit acting as a probe. We highlight, like an ideal quantum simulator, the system can be simulated at any arbitrary point on its \emph{amoeba} at any arbitrary temperature. Even the internal coupling of the Ising system can be set to a desired value ranging from ferromagnetic to anti-ferromagnetic regimes. This initialization is accomplished solely by manipulating the probe qubit, leaving the system qubits undisturbed. After effectively initiating the system at an arbitrarily chosen point on its \emph{amoeba}, corresponding points of \emph{coamoeba} can be directly sampled from the time evolution of the probe's coherence. By iterating the process, one samples the \emph{coamoeba} across the \emph{amoeba} to obtain the full \emph{algebraic variety}. Thus both preparation and detection are achieved through the probe alone. We demonstrate the method experimentally via a three-qubit NMR register by taking two of them as system and the third one as probe.  Sampling of \emph{coamoeba} at different points on \emph{amoeba} is performed directly from time domain NMR signal without any need for extensive post-processing of experimental data.  Apart from extending the range of quantum simulations to other areas of physics where determining LY zeros of general classical system is pivotal, our work also brings pure mathematical structures like \emph{amoeba} and \emph{coamoeba} into the realm of quantum simulations by physically sampling them for a given LY polynomial.

The rest of the article is organized as follows. In Sec.~\ref{sec:Ising}, we introduce the asymmetric Ising system and show how to use qubits as system and probe such that the \emph{amoeba} and the \emph{coamoeba} of the \emph{algebraic variety} corresponding to the LY polynomial of the system relate to probe qubit's coherence. After describing the methodology of sampling the \emph{algebraic variety} in Sec.~\ref{sec:Method}, we discuss how to initialize the system qubits at any desired point on the \emph{amoeba} at any value of physical parameters like temperature and coupling constant by operating on only the probe in Sec.~\ref{sec:Initialization}. Finally, we present the experimental results in Sec.~\ref{sec:Results} before concluding in Sec.~\ref{sec:Conclusion}.  

\begin{figure}
	\centering 
	\includegraphics[scale=0.16]{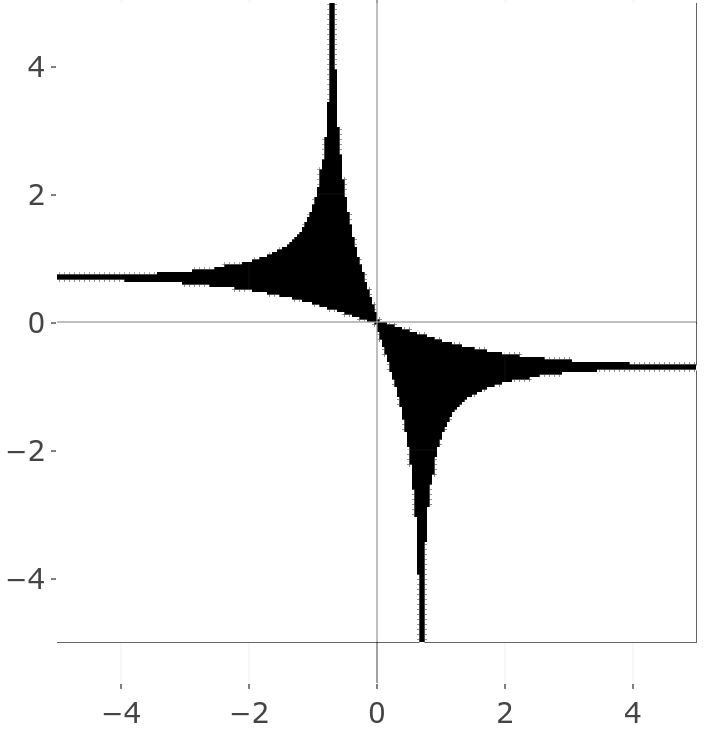}
	\includegraphics[scale=0.75]{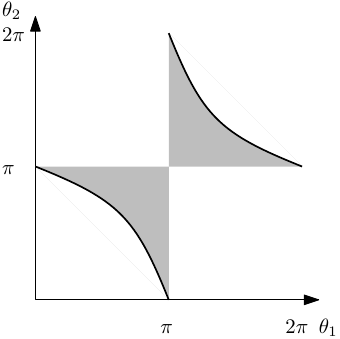} \\
	(a) \hspace{3.5cm} (b)
	\caption{The \emph{amoeba} (a) and \emph{coamoeba} (b) of the LY polynomial $f(z_1,z_2)=1+\half z_1+\half z_2+z_1z_2$~ \cite{sadyT}.}
	\label{fig:example}
\end{figure}	

\section{Two-spin Ising model and (co)amoebas} \label{sec:Ising}
To explain the method, we consider our system to be a classical Ising chain consisting of only two sites, $A$ and $B$, coupled to each other with strength $J$. Without assuming any symmetry, let the magnetic fields at sites $A$ and $B$ be $h_A$ and $h_B$, respectively. This classical system can be mimicked by using two spin $1/2$ systems as qubits with Hamiltonian $H_{AB}=-J\sigma_z^A\sigma_z^B-h_A\sigma^A_z-h_B\sigma^B_z$, where $\sigma_i^X$ is the Pauli $i$-operator of spin $X$.

In a thermal bath of inverse temperature $\beta$, the partition function of $AB$ is 
\begin{align}
Z(\beta,h_A,h_B)=&\expo{\beta J+\beta h_A+\beta h_B}\big(1+\expo{-2\beta J-2\beta h_A}+
\nonumber \\
&~~~~~\expo{-2\beta J-2\beta h_B}+\expo{-2\beta h_A-2\beta h_B}\big).
\end{align}
We extend $Z$ into the complex domain by letting 
\begin{align}
	\expo{-2\beta h_A}=z_1, \expo{-2\beta h_B}=z_2; ~ \mathrm{where} ~~  z_1,z_2 \in (\mathbb{C}^*)^2. \label{eq:complex_extension}
\end{align}
Further, by identifying $\Gamma =\expo{-2\beta J}\in\mathbb{R}_+$, the partition function becomes $Z=(\Gamma z_1z_2)^{-1/2}f(z_1,z_2)$, where 
\begin{align}
	f(z_1,z_2)=1+\Gamma z_1+\Gamma z_2+z_1z_2 \label{eq:LYpolynomial}
\end{align}
is the two-spin bivariate \emph{LY polynomial}. Zeros of the partition function now correspond to the vanishing of this polynomial, which defines the \emph{algebraic variety} $V_f=\left\{(z_1,z_2)\in\brac{\mathbb{C}^*}^2\;\big|\; f(z_1,z_2)=0 \right\}$.

For any complex number $(z_1,z_2)$ lying on $V_f$, information about its absolute value is studied using a log map $\mathrm{Log}:(\mathbb{C}^*)^2\rightarrow\mathbb{R}^2$ by taking $(z_1,z_2)\mapsto(\log|z_1|,\log|z_2|)$. The image $\mathcal{A}_f=\mathrm{Log}(V_f)$ is called \emph{amoeba}. On the other hand, information about the phase is studied using the argument map $\mathrm{Arg}:(\mathbb{C}^*)^2\rightarrow S^1\times S^1$ where $(z_1,z_2)\mapsto(\arg {z_1},{\arg z_2})$. The image $\mathrm{co}\mathcal{A}_f=\mathrm{Arg}(V_f)$ is called the \emph{coamoeba}. For example, the \emph{amoeba} and \emph{coamoeba} for $\Gamma=\half$ in Eq.\Eqref{eq:LYpolynomial} are shown in Fig.~\ref{fig:example}.

\begin{figure}
	\centering
	\includegraphics[height = 3cm, trim = {0cm, 10.5cm, 2.21cm, 2cm},clip]{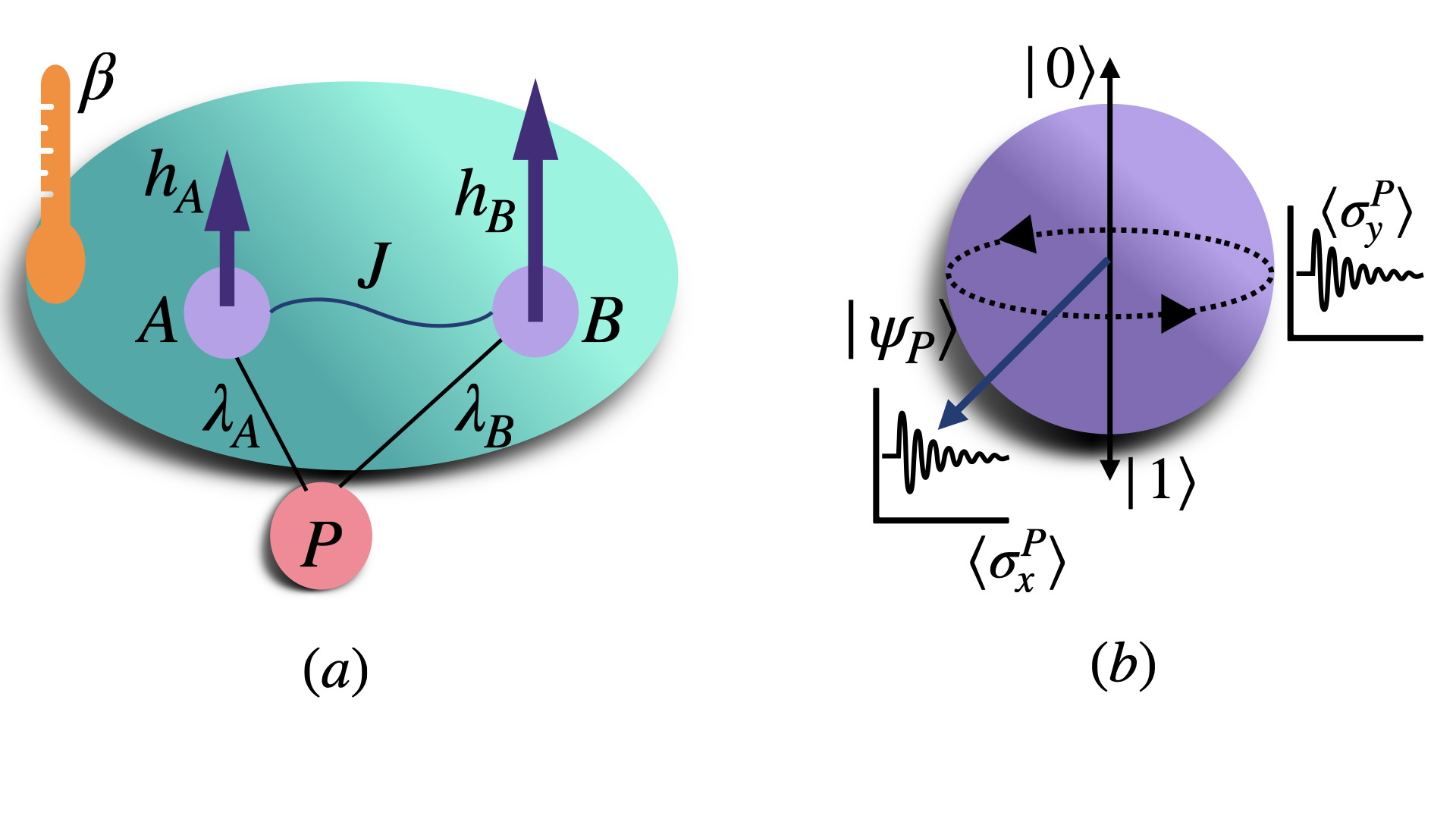} \\
	(a)  \hspace{4cm}  (b)
	\caption{(a) A two-spin Ising system in a thermal bath at inverse temperature $1/\beta$ coupled to a probe spin.  (b) As the probe evolves under the interaction with the system, it's coherence is recorded by measuring $\langle \sigma_x^P \rangle $ and $\langle \sigma_y^P \rangle$ with time.}
	\label{fig_SpinBath}
\end{figure}

To observe these complex zeros, we adopt the procedure of Wei and Liu \cite{PhysRevLett.109.185701} by introducing another spin-$1/2$ particle, which acts as a quantum probe. Its states $\{\ket{0},\ket{1}\}$ (here $\ket{0}$ and $\ket{1}$ are eigenvectors of $\sigma_z$ with eigenvalues $\pm 1$, respectively) span a two-dimensional Hilbert space $\mathcal{H}_{P}$. The total space is now $\mathcal{H}_{\mathrm{tot}}=\mathcal{H}_{P}\otimes\mathcal{H}_{AB}$. However, here we allow the probe to couple asymmetrically with $A$ and $B$. This is because, in our case, $h_A$ and $h_B$ are not necessarily equal, and hence $z_1$ and $z_2$ must be distinct variables, necessitating the description of Lee-Yang zeros in $(\mathbb{C}^*)^2$ rather than in $\mathbb{C}$.

Suppose that the probe spin $\sigma^P_z$ couples to $\sigma_z^A$ and $\sigma_z^B$ with coupling strengths $\lambda_A$ and $\lambda_B$, respectively. This situation is depicted in Fig.~\ref{fig_SpinBath}~a. The interaction Hamiltonian is {}
\begin{align}
	H_{\mathrm{int}}=\lambda_A\sigma^P_z\sigma_z^A + \lambda_B\sigma^P_z\sigma_z^B. \label{eq:Hint}
\end{align}
Here we used the notation $\sigma^P_z\sigma^A_z=\sigma^P_z\otimes\sigma_z^A\otimes\mathbbm{1}$ and $\sigma^P_z\sigma^B_z=\sigma^P_z\otimes\mathbbm{1}\otimes\sigma_z^B$. The full Hamiltonian now becomes $H=\mathbbm{1}\otimes H_{AB}+H_{\mathrm{int}}$, or 
\begin{equation}
	H=-J\sigma^A_z\sigma^B_z-(h_A-\lambda_A\sigma^P_z)\sigma^A_z-(h_B-\lambda_B\sigma^P_z)\sigma_z^B. \label{eq:Hth}
\end{equation}
The full initial state at time $t=0$ is taken to be
\begin{align}
	\rho(0)=\ket{\psi_P}\bra{\psi_P}\otimes\frac{\exp\brac{-\beta H_{AB}}}{Z(\beta,h_A,h_B)},\label{eq:initial-state}
\end{align}
where $\ket{\psi_P}$ is a pure coherent quantum state in the probe's $\sigma_z$ basis. Here, by a coherent state we mean that $\ket{\psi_P}$ has superposition in $\{\ket{0},\ket{1}\}$ basis. It subsequently evolves under the Hamiltonian in Eq.~\Eqref{eq:Hth} as $\rho(t) = \mathcal{U}(t)\rho(0)\mathcal{U}^{\dagger}(t)$, where $\mathcal{U}(t) = \expo{-\im t H/\hbar}$. As we only observe the evolution of the probe, we trace out $AB$ to get $\rho_{P}(t)=\mathrm{tr}_{AB}\rho(t)$.
Under this evolution, the coherence $L$ of probe at time $t$ can be shown to take the form 
\begin{align}
	L(t)&= \frac{\hbar^2}{4}\left|\langle\sigma_x^P(t)\rangle+\im\langle\sigma_y^P(t)\rangle \right|^2  = \mathcal{C}\left|f(z_1,z_2)\right|^2, \label{eq:Lt}
\end{align}
derivation can be found in Appendix~\ref{appenA}. Here  $\mathcal{C}$ is a constant over time and $f(z_1,z_2)$ is the two-spin  bivariate LY polynomial of Eq.~\Eqref{eq:LYpolynomial}, after identifying
\begin{align}
z_1&=\exp\brac{-2\beta h_A+4\im\lambda_At/\hbar},
\nonumber \\
z_2&=\exp\brac{-2\beta h_B+4\im\lambda_Bt/\hbar}, ~\mbox{and} \nonumber \\
\Gamma&=\exp\brac{-2\beta J}.
\end{align}
Note that the variables $z_1$ and $z_2$ become complex as the time evolution introduces a complex phase. Furthermore, we note 
\begin{subequations}
	\begin{align}
		\log|z_1|&=-2\beta h_A,\hspace{1.05cm} \log|z_2|=-2\beta h_B, ~~\mathrm{and} \label{eq:am}\\
		\theta_1(t)&=\arg z_1=\frac{4\lambda_A t}{\hbar},\quad \theta_2(t)=\arg z_2=\frac{4\lambda_B t}{\hbar}. \label{eq:coam}
	\end{align}
\end{subequations}
Corresponding to the set of points $\{(z_1,z_2)\}$ for which $|f(z_1,z_2)|^2=0$, the set $\{(\log|z_1|,\log|z_2|)\}$  forms the \emph{amoeba} while the set $\{(\arg z_1,\arg z_2)\}$ forms the \emph{coamoeba}.

\section{Method for Observing The Algebraic Variety}  \label{sec:Method}
Our method for determining the \emph{algebraic variety} $V_f$ works as follows: we first initiate the system in the state given by  Eq.~\Eqref{eq:initial-state} at arbitrary values of $\beta h_A$ and $\beta  h_B$, which according to Eq.~\Eqref{eq:am}, fix a point on the \emph{amoeba} space. As we shall see later, the preparation is achieved through the probe, assuming no control over the system qubits. Next, we let the probe interact with the system and observe its coherence. If the coherence is non-zero at all finite times, we conclude that there are no LY zeros at that value of $(\beta h_A,\beta h_B)$; hence, the chosen point does not belong to the \emph{amoeba}. On the other hand, if we find the coherence vanishing at time instants $\{t_i\}$, the point $(\beta h_A,\beta h_B) \equiv (\log|z_1|,\log|z_2|)$ belongs to the \emph{amoeba}. Moreover, by Eq.~\Eqref{eq:coam},  $\{t_i\}$ can be mapped to $\{(\theta_1(t_i),\theta_2(t_i))\}$, which lie on a $S^1\times S^1$ torus as the points of \emph{coamoeba}, corresponding to the point  $(\log|z_1|,\log|z_2|)$ on \emph{amoeba}. Upon multiple iterations of this procedure, one can sample the \emph{coamoeba} across the \emph{amoeba} to extract the full $V_f$ for the multivariate LY polynomial. For more details regarding the sampling of \emph{coamoeba}, see Appendix~\ref{AppenB}.

Another way of seeing the fact that complex LY zeros leave footprint in the real time dynamics of the probe would be in terms of correlation. Mutual information \cite{nielsen2010quantum} between probe and the system qubits captures the total correlation between them and is defined as 
\begin{align}
I_{P:AB}(t) = S_{P}(t) + S_{AB}(t) - S_{PAB}(t),   \label{eq:MI}
\end{align}
where $S_{X}(t) = -\mathrm{tr}\left[\rho_X(t) \log \rho_X(t)\right]$ is the Von-Neumann entropy \cite{nielsen2010quantum} of  $X$ at time $t$. By evolving the initial state of Eq.~\Eqref{eq:initial-state} under the Hamiltonian of Eq.~\Eqref{eq:Hth}, it can be seen that $S_{AB}$ and $S_{PAB}$ do not evolve over time. Therefore, according to Eq.~\Eqref{eq:MI}, $I_{P:AB}$ becomes maximum at times only when the probe is maximally mixed, i.e $\rho_{P} = \mathbbm{1}/2$. The reduced density matrix of the probe can be expanded in Pauli basis as
\begin{align}
\rho_P(t) = \frac{1}{2} \left(\mathbbm{1} + \sum_{i \in \{x,y,z\}} c_i(t) \sigma_{i}^P \right), \label{eq:MIstate}
\end{align}
where $c_i (t)= \langle \sigma^P_i (t) \rangle$.  It can be easily seen that $\partial_{t}c_z=0$ in the evoltion under the Hamiltonian of Eq.~\Eqref{eq:Hth} . If the intial state of the probe is such that $c_z(0)=0$, then its reduced density matrix becomes maximally mixed at time points where $\langle \sigma_x^P \rangle$ and $\langle \sigma_y^P \rangle$ simultaniously vanish, i.e points corresponding to LY zeros. Therefore, only at the time points corresponding to LY zeros, the correlation between the probe and the system becomes maximum.

\section{System Initialization via a Quantum Probe} \label{sec:Initialization}
\begin{figure}
	\centering
	\includegraphics[ width=6.7cm, trim={0cm 4cm 0cm 0cm},  clip=true]{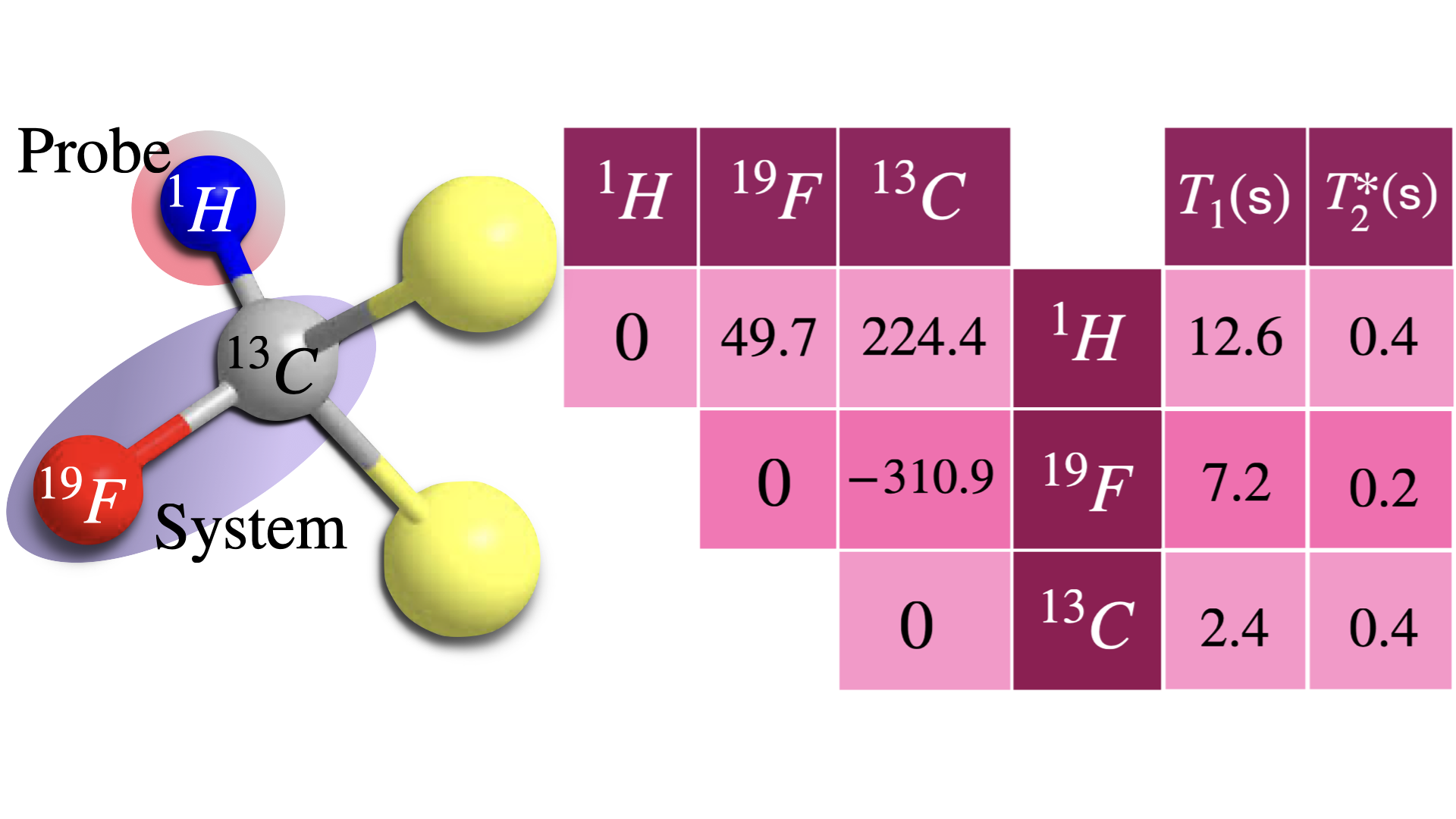} \\
	(a) \hspace{3.3cm} (b) 
	\caption{(a) The molecular structure of DBFM with labeled spins. (b) Values of NMR Hamiltonian parameters (resonance offsets in diagonal and coupling constants $J_{ij}$ in off-diagonal elements, in Hz) along with relaxation time constants.} 
	\label{fig:molecule}
\end{figure}

\subsection{NMR register}
For a concrete demonstration via NMR,  we use a liquid ensemble of three spin-1/2 (three-qubit) nuclear register of $^{13}C$-Dibromofluoromethane (DBFM) (see Fig.~\ref{fig:molecule}~(a)), dissolved in Acetone-D6, by identifying $^{1}H$ as probe $P$ and $^{19}F$, $^{13}C$ as system spins $A, B,~$respectively. In a high static magnetic field of $B_0 = 11.7$~T, their Larmor frequencies have magnitudes $\omega_i = \gamma_{i}B_0$, where $\gamma_i$ are the gyromagnetic ratios \cite{cavanagh1996protein}.  The three spins also interact mutually via scalar coupling with strengths $J_{ij}$ as tabled in Fig.~\ref{fig:molecule}~(b). In terms of the spin operators $I_\alpha^i = \sigma_\alpha^i/2$, the lab-frame NMR Hamiltonian is of the form $H_{\mathrm{NMR}} = H_I + H_{RF}$, with internal part $H_{I} = -\sum_{i} \hbar \omega_{i} I_z^i + 2\pi \hbar \sum_{i \neq j} J_{ij} I_z^i I_z^j$  
and the probe control part $H_{RF}=-\hbar \Omega_{P}(t) I_{x}^{P} $. 
Here $\Omega_{P}(t) = \gamma_i B_P(t)$  represents the control amplitude achieved through the magnetic component $B_P(t)$ of the applied circularly polarized RF field resonant with the probe's Larmor frequency. Notice, the control Hamiltonian contains only the probe's component $\Omega_{P}(t)$, while we set $\Omega_{A}(t)=\Omega_{B}(t)=0$ to ensure that the method works without assuming any experimental control over the system.

We allow the liquid ensemble register to equilibrate at an ambient temperature of $T = 300$ K inside a $500$ MHz Bruker NMR spectrometer.
As $\hbar \omega_i \ll k_{B}T$, the density operator in thermal state becomes  $\rho_{\text{th}}  = \exp(-\beta H_I)/ \mathrm{tr} [\exp(-\beta H_I)] \approx \mathbbm{1}/8 + \epsilon\rho_{\text{th}}^\Delta$, where $\rho_{\text{th}}^\Delta =  I_{z}^{H} + (\gamma_{F}/\gamma_{H}) I_{z}^{F} + (\gamma_{C}/\gamma_{H}) I_{z}^{C}$ is the deviation thermal state and 
the purity factor $\epsilon \approx 10^{-5}$.  From the thermal state,  preparation of the initial state in Eq.~\Eqref{eq:initial-state}  at any chosen values of $\left\{\beta h_A, \beta h_B\right\}$ is to be achieved, subject to constraint that the system ($^{19}F,~^{13}C$) remains inaccessible. From now onward the probe's state of Eq.~\Eqref{eq:initial-state} is taken to be $\ket{\psi_P} = (\ket{0} + \ket{1})/\sqrt{2}$  

\begin{figure*}
\centering
\includegraphics[ width=15cm, trim={12cm 0cm 14cm 0cm},  clip=true]{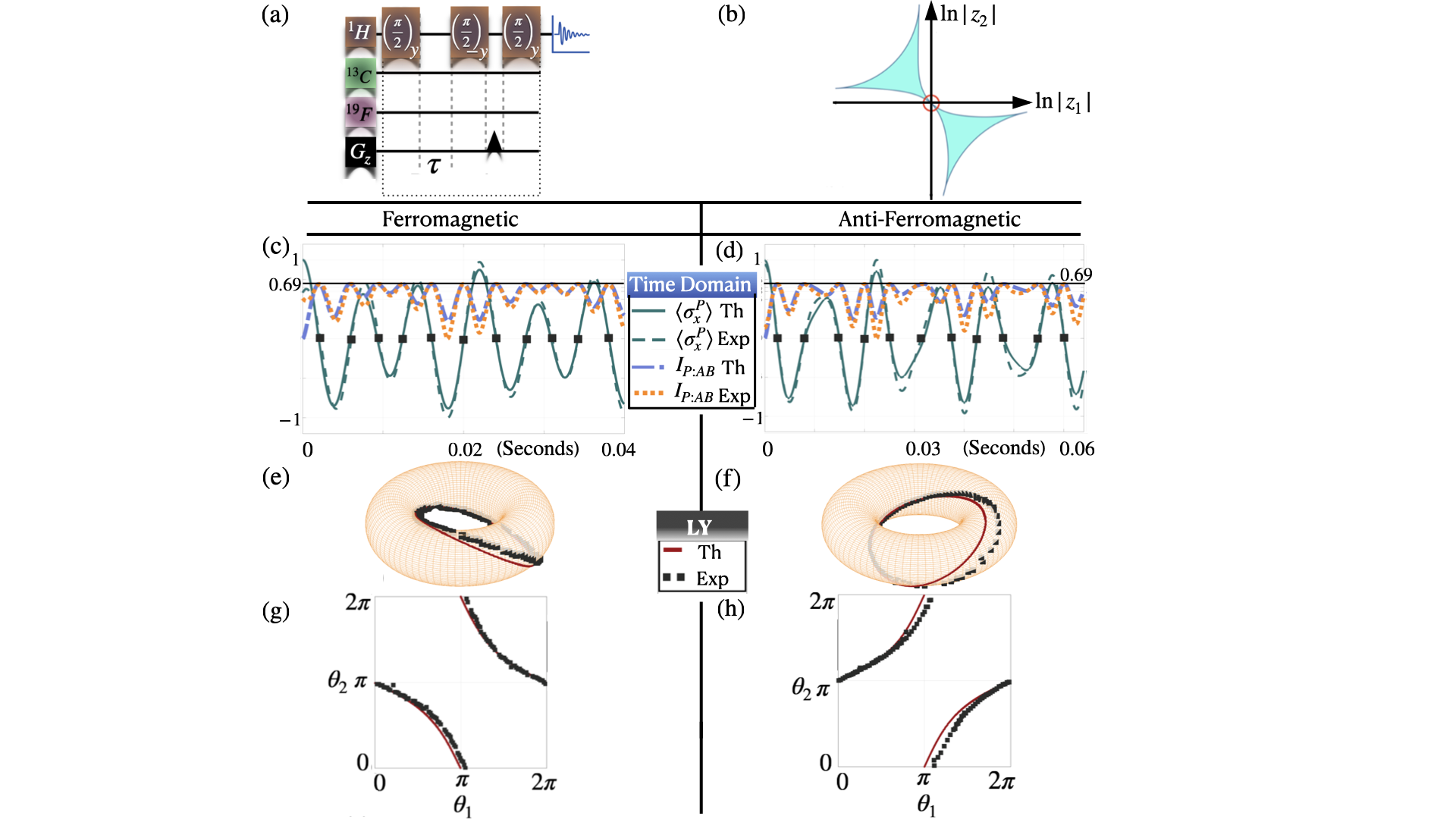} \\
\caption{Pulse sequence (a) for preparing the system in state of Eq.~\Eqref{eq:initial-state} for $h_A = h_B = 0$. We set $\tau = 8.76$ ms or $\tau =9.013$ ms to prepare $\beta J= 0.5$ (ferromagnetic case) or $\beta J=-0.5$ (anti-ferromagnetic case) respectively.  The corresponding \emph{amoeba} point is at the origin of the complex plane as illustrated in (b).  (c-d) The real parts of FIDs (solid lines) corresponding to $\langle \sigma_x^P(t) \rangle$  are plotted along with the simulated curves (dashed lines) for (c) ferromagnetic and (d) anti-ferromagnetic cases respectively. As $\langle \sigma_y^P (t) \rangle$ is identically zero in these cases, the FID null points (solid squares) correspond to LY zeros. Theoretical and experimental value of  the mutual information $I_{P:AB}$ between the probe and the system are also plotted, which reach their maxima (horizontal line at $0.69$) only at the LY zeros. (e-f) Theoretically calculated (solid line) and experimentally observed (filled squares) \emph{coamoeba} on a 2-torus for (e) ferromagnetic and (f) anti-ferromagnetic cases respectively.  The corresponding planar visualizations are in (g-h). }
\label{fig:ImExp}
\end{figure*}

\subsection{Initializing at the Origin of Amoeba Plane}

First consider the preparation for $h_A = h_B = 0$, which corresponds to the origin of \emph{amoeba} in accordance to Eq.~\Eqref{eq:am}. We note that the factor $\exp(-\beta H_{AB})$ of Eq.~\Eqref{eq:initial-state} can be expanded as 
\begin{align}
\expo{-\beta H_{AB}}  = &\expo{\beta J \sigma_z^A \sigma_z^B} \expo{\beta h_A \sigma_z^A} \expo{\beta h_B \sigma_z^B} \nonumber \\
 = C_a \mathbbm{1}_4 + & C_b \sigma_z^A + C_c \sigma_z^B + C_d \sigma_z^A  \sigma_z^B, \nonumber \\
\mathrm{where,} ~~ C_a &= [\cosh(\beta J) \cosh(\beta h_A) \cosh(\beta h_B) \nonumber \\
&+ \sinh(\beta J) \sinh(\beta h_A) \sinh(\beta h_B)], \nonumber \\
C_b &=  [\cosh(\beta J) \sinh(\beta h_A) \cosh(\beta h_B) \nonumber\\
&+ \sinh(\beta J) \cosh(\beta h_A) \sinh(\beta h_B)],  \nonumber \\
C_c &=  [\cosh(\beta J) \cosh(\beta h_A) \sinh(\beta h_B) \nonumber\\
&+ \sinh(\beta J) \sinh(\beta h_A) \cosh(\beta h_B)],  \nonumber \\
C_d &=  [\cosh(\beta J) \sinh(\beta h_A) \sinh(\beta h_B) \nonumber\\
&+ \sinh(\beta J) \cosh(\beta h_A) \cosh(\beta h_B)]  \label{eq:series}. 
\end{align}
Substituting the above to Eq.~\Eqref{eq:initial-state} and setting $h_A=h_B=0$, the initial state in this case becomes $\rho(0) = \mathbbm{1}/8 + (1/2) \tanh (\beta J) I_z^A I_z^B +  \left[\cosh(\beta J) I_x^P + \sinh (\beta J) 4 I_x^P I_z^A I_z^B\right]/Z(\beta,0,0)$. Here the first two terms can be suppressed as they remain invariant under time evolution governed by $H_I$, and also do not contribute to probe's coherence in Eq.~\Eqref{eq:Lt}. Therefore, the target state for initialization reads
\begin{equation}
\rho^{\Delta}(0) =  \left[\cosh(\beta J) I_x^P + \sinh (\beta J) 4 I_x^P I_z^A I_z^B\right]/ Z(\beta,0,0),	\label{eq:targIm}
\end{equation}
where $\Delta$ is used as superscript to indicate that redundant terms are suppressed as mentioned earlier.  For the same reason we ignore the identity term of $\rho_{\text{th}}$ and consider ony the deviation thermal state  $\rho^{\Delta}_{\mathrm{th}}$. We have to realize the target initial state $\rho^{\Delta}(0)$ of Eq.~\Eqref{eq:targIm}  from the deviation thermal state $\rho_\mathrm{th}^\Delta$. In other words, for initialization, the target state of Eq.~\Eqref{eq:targIm} is to be achieved starting from $\rho_\mathrm{th}^\Delta(0)$ with the constrain that the system remains inaccessible. For this purpose, we employ the pulse sequence shown in Fig.~\ref{fig:ImExp}~(a) which operates on probe spin of $^{1}H$ only. Detailed derivation of the pulse sequence is given in Appendix~\ref{AppendC1}. Here, the delay time $\tau$ of the pulse sequence is a free parameter that depends directly on the value of $\beta J \in \mathbb{R}$  of the target state as $\exp(\beta J)=\cos(\pi(J_{PA}+J_{PB})\tau)/\cos(\pi(J_{PA}-J_{PB})\tau)$. This direct dependency  allows us to initiate the system at any value of $\beta J$ in both ferromagnetic ($\beta J \in \mathbb{R}_{+}$) and anti-ferromagnetic ($\beta J \in \mathbb{R}_{-}$) regime just by varying the delay $\tau$ suitably, as discussed in detail in Appendix~\ref{AppendC1}.

\subsection{Initialization on non-origin points of Amoeba Plane}

To sample \emph{coamoeba} corresponding to non-origin points on \emph{amoeba}, we need $h_A$ and $h_B$ to take non-zero values. The state of Eq.~\eqref{eq:initial-state} can be directly computed as $\rho(0) =  (\mathbbm{1}/2 + I_x^P) \otimes (\expo{-\beta H_{AB}}/Z(\beta J,h_A,h_B))$, which, upon substituting the value of $\expo{-\beta H_{AB}}$ from Eq.~\Eqref{eq:series} yields 
\begin{equation}
\rho^\Delta (0) = C_a I_x^P + C_b 2I_x^P I_z^A + C_c 2I_x^P I_z^B + C_d 4I_x^P I_z^A I_z^B.  \label{eq:target}
\end{equation}	
Here, we have used `tilde' as before to indicate that only those terms that contribute to the probe's coherence are considered whereas $\mathbbm{1}$ and $I_z^A I_z^B$ are suppressed.
$C_i$'s of Eq.~\Eqref{eq:target} are hyperbolic functions of $\beta h_A$, $\beta h_B$ and $\beta J$ as defined in Eq.~\Eqref{eq:series}. Again, starting from the thermal deviation $\rho_{\text{th}}^\Delta$, target state $\rho^\Delta (0)$ of Eq.~\Eqref{eq:target} can be prepared with only probe control using the pulse sequence shown in Fig.~\ref{fig:CompExp}~(a). The derivation of pulse sequence is given in Appendix~\ref{AppendD1} which explains how the pulse sequence achieves the desired initialization. As there are three variables $\left( \beta J,~\beta h_A,~\beta h_B\right)$  to fix the target state of Eq.~\Eqref{eq:target}, the pulse sequence of Fig.~\ref{fig:CompExp}~(a) is also having three free parameters $\theta^1,~\theta^2$ and $\tau$, which can be set suitably to prepare any desired state. Given a specific target state, how to obtain the correct values of three control parameters $\theta^1,~\theta^2$ and $\tau$ is discussed in Appendix~\ref{AppendD1}.

\begin{figure*}
	\centering
	\includegraphics[width=15cm, trim={12.2cm 0cm 13.9cm 0cm},  clip=true]{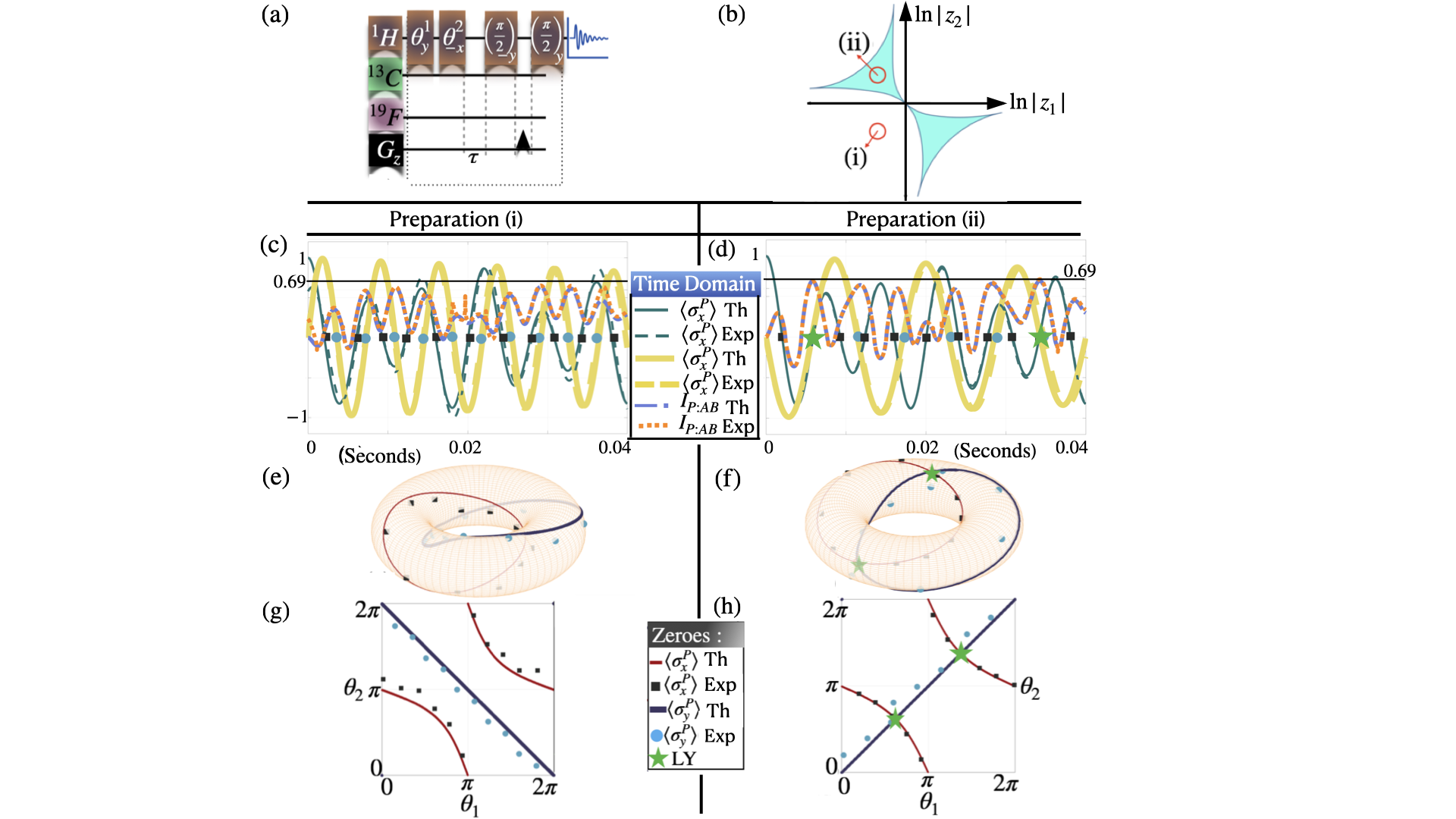} \\
	\caption{
		Pulse sequence (a) for preparing the system in state of Eq.~\Eqref{eq:initial-state} for non zero values of $h_A$ and $h_B$. In particular, the initialization is performed for two distinct points on \emph{amoeba} plane (b) :(i) $h_A=h_B=0.1$ and (ii) $h_A=-h_B=0.1$, considering $\beta J = 0.5$ in both cases.
		(c-d) The real parts (solid thin lines) and imaginay parts (solid thick lines) of FID corresponding to $\langle \sigma_x^P(t) \rangle$  and  $\langle \sigma_y^P(t) \rangle$, respectively,  are plotted along with the simulated curves (dashed lines, thin and thick correspondingly) for prepared initial state of (i) in (c) and (ii) in (d). Simultaneous null points of real (solid square) and imaginary part (solid circle) of FID correspond to LY zeros, which exists for (ii) (stars) while absent for (i). Theoretical and experimental curves of the mutual information $I_{P:AB}$ between the probe and the system are also plotted, which reach their maxima (horizontal line at $0.69$) only at the LY zeros. (e-f) Theoretically calculated (solid line) and experimentally observed null points of real (solid square) and imaginary (solid circle) part of FID are plotted on the 2-torus for (i)(e) and (ii)(f). The desired \emph{coamoeba} points lie in the intersetion, which can be seen in (f). The corresponding planar visualizations are in (g-h). }
	\label{fig:CompExp}
\end{figure*}

\section{Experimental Results} \label{sec:Results}

As explained above, LY zeros are extracted from the time points where probe coherence vanishes, which, according to Eq.~\Eqref{eq:Lt}, leads to simultaneous vanishing of $\langle \sigma_{x}^P\rangle$ and $\langle \sigma_{y}^P\rangle$.  As the prepared state $\rho^\Delta (0)$ evolves freely under the internal Hamiltonian $H_I$, we measure the probe expectation values $\langle \sigma_{x}^P(t)\rangle$ and $\langle \sigma_{y}^P(t)\rangle $ in a rotating frame synchronous with the probe's Larmor precession \cite{cavanagh1996protein,levitt2013spin}. In this frame, the effective Hamiltonian becomes
\begin{align}
	H_{\mathrm{eff}} = 2\pi \hbar \left( J_{PA} I_z^{P} I_{z}^{A} + J_{PB} I_z^{P} I_{z}^{B}  \right),  \label{eq:Heff}	
\end{align}
where $A$-$B$ interaction is suppressed since $\rho^\Delta(0)$ of Eq.~\Eqref{eq:targIm} and \Eqref{eq:target} do not evolve under it.  By comparing this Hamiltonian with the interaction Hamiltonian $H_{\mathrm{int}}$ in Eq.~\Eqref{eq:Hint}, we identify $\lambda_A = \pi J_{PA} \hbar /2$ and $\lambda_B = \pi J_{PB} \hbar /2$.  Feeding this to Eq.~\Eqref{eq:coam}, experimentally measured time points corresponding to vanishing probe coherence are directly mapped to the \emph{coamoeba} torus, and thus, physical sampling of \emph{coamoeba} is achieved. 

First, we sample the \emph{coamoeba} corresponding to the origin (circled in Fig.~\ref{fig:ImExp}~(b)) of the \emph{amoeba}'s plane. Following Eq.~\Eqref{eq:am}, we set $h_A = h_B = 0$ and accordingly prepare $\rho^\Delta(0)$ of Eq.~\Eqref{eq:targIm}, which subsequently evolves under $H_\mathrm{eff}$.  As $\langle \sigma^P_y(t) \rangle$ is identically zero in this case (see Appendix~\ref{AppendC1} for calculation),  LY zeros are determined just by measuring $\langle \sigma_x^P (t) \rangle$ and noting the time points where it vanishes. 
An essential advantage of the NMR quantum testbed is, being an ensemble architecture, it directly gives $\langle \sigma^P_{x(y)}  (t)\rangle$ as the real (imaginary) part of the NMR signal known as the free induction decay (FID) \cite{fukushima2018experimental}. 
Hence, after performing the initialization via a probe in ferromagnetic and anti-ferromagnetic regimes, we record their FIDs, whose real parts are shown in Fig.~\ref{fig:ImExp}~(c) and (d), respectively. 
In the figure, time points corresponding to LY zeros are marked with solid squares, which are readily identified as FID null points. No additional data processing is required for this purpose. 
We sampled the corresponding \emph{coamoeba(s)} from collected FID up to $300$ ms (see Appendix~\ref{AppendC2}). However, for clarity, the time domain signal for ferromagnetic and anti-ferromagnetic cases in Fig.~\ref{fig:ImExp}~(c) and (d) are shown only up to $40$ and $60$ ms, respectively. The mutual information $I_{P:AB}$ between the probe and the system is calculated directly from experimental FID and are plotted against their simulated values in Fig.~\ref{fig:ImExp} (c,d). It is shown that the maximum of the mutual information occurs only at the LY zeros as predicted by Eq.~\Eqref{eq:MI}, thus confirming the footprint of complex LY zeros in real-time correlation dynamics between the system and the probe.
The FID null points are mapped to the \emph{coamoeba} by Eq.~\Eqref{eq:coam}. The sampled \emph{coamoeba} for ferromagnetic ($\beta J= 0.5$) and anti-ferromagetic ($\beta J = -0.5$) regimes are shown on the $2$-torus in Fig.~\ref{fig:ImExp}~(e) and (f)  respectively.  Their topologically equivalent modulo $2\pi$ planar visualizations are shown in Fig. ~\ref{fig:ImExp}~(g) and (h), respectively. We observe a fairly good agreement with the theoretically computed \emph{coamoeba} within the experimental limitations.

It is worth noting that no state tomography is needed for these experiments, and the LY zeros emerge directly from the NMR FID without further data processing. Thus, each experiment taking less than a second yields a dense set of LY zeros.

We now demonstrate the sampling of \emph{coamoeba} for two non-origin points, (i) $\beta h_A = \beta h_B = 0.1$, which is outside \emph{amoeba}, and (ii) $\beta h_A = -\beta h_B = 0.1$, which is inside \emph{amoeba} as shown in Fig.~\ref{fig:CompExp}~(b).  Again we note that $\langle \sigma_x^P (t) \rangle$ and $\langle \sigma_y^P (t) \rangle$ are just real and imaginary components of the probe ($^1H$) FID. Therefore in each case, we initialize the system considering $\beta J=0.5$, let it evolve under $H_\mathrm{eff}$ while recording the FID as shown in Fig.~\ref{fig:CompExp}~(c-d).We extract the time points at which $\langle \sigma_x^P(t) \rangle$ and $\langle \sigma_y^P(t) \rangle$ vanish, and map them to a 2-torus via Eq.~\Eqref{eq:coam}. In case (i), as shown in Fig.~\ref{fig:CompExp}~(c,e,g), the zeros of $\langle \sigma_x^P(t) \rangle$ and $\langle \sigma_y^P(t) \rangle$ do not intersect indicating the absence of L-Y zeros, thereby confirming that the point (i)  does not belong to the \emph{amoeba}.  
However in case (ii), the null points of $\langle \sigma_x^P(t) \rangle$ (real FID) and $\langle \sigma_y^P(t) \rangle$ (imaginary FID) intersect twice as marked by stars in Fig.~\ref{fig:CompExp} (d,f,h), confirming the existence of two distinct \emph{coamoeba} points. The mutual information calculated from the NMR FID is plotted along with its simulated values for case (i) [(ii)] in Fig.~\ref{fig:CompExp} (c) [(d)]. It reaches its maximum twice for case (ii) at times corresponding to simultaneous null points of real and imaginary components of the FID. However, as there are no simultaneous null points of real and imaginary FID in case (i), the mutual information never becomes maximum in this case. These experimental observations confirm the prediction of Eq.~\Eqref{eq:MI} that the correlation between the probe and the system reaches its maximum only at points corresponding to LY zeros. It is worth highlighting that the existence and non-existence of LY zeros for case (i) and (ii) , respectively, can be directly observed just by looking at quadrature NMR FID shown in Fig.~\ref{fig:CompExp}~(c-d) without any data processing.
Again, in both cases, we see a reasonably good agreement between the theoretical predictions and experimental values. This method of high-throughput extraction of LY zeros can be used for efficient sampling of \emph{coamoeba} for a large set of \emph{amoeba} points, thereby determining the \emph{algebraic variety} $V_f$ at any desired precision. 

\section{Conclusion}  \label{sec:Conclusion}
For the continued advancement of quantum technologies in the coming years, it is imperative that their applications extend to a broader spectrum of scientific domains by addressing challenges beyond the confines of problems related to quantum physics alone. Following the spirit, we showed a method of using qubits to simulate asymmetrical classical Ising systems at any arbitrary value of its temperature and coupling constant for determining its LY zeros in a wide range of physical situations. Most importantly, in our method, both initialization and determination of LY zeros are achieved through a quantum probe interacting with the system qubits while system qubits themselves left untouched. We believe, this feature of our protocol makes it easier to generalise for more complex systems where controlling system qubits become intractable. A lot of recent works presented a wide range of applications of LY zeros in solving problems across areas like statistical studies  \cite{PhysRevE.97.012115, PhysRevC.72.011901} of equilibrium (phase transition, critical phenomena etc \cite{doi:10.1126/sciadv.abf2447, PhysRevResearch.1.023004}) and non-equilibrium (dynamical phase transitions etc \cite{PhysRevLett.118.180601}) statistical physics, percolation   \cite{arndt2001directed}, complex networks \cite{krasnytska2015violation,krasnytska2016partition} and even protein folding \cite{PhysRevLett.110.248101,PhysRevE.88.022710}. Therefore a method for extracting full \emph{algebraic variety} containing LY zeros of a general asymmetrical classical system has become essential to implement these studies in real situations. We believe, our method of using quantum simulation technique with control over a single qubit alone to do the task is a pioneering step in bringing all those different areas of physics to the sphere of quantum simulation. Experimental validation using a three-qubit NMR register demonstrates the feasibility of our protocol. It is worth noting that this protocol samples the \emph{amoeba} and \emph{coamoeba} for a given LY polynomial and thereby can provoke applications of quantum simulations in the domain of pure mathematics. Apart from applications, this work also uncovers the rich aesthetic structure of the LY zeros by physically sampling the \emph{algebraic variety} that contains them.

\section{Acknowledgments}
AC acknowledges K. Arora for discussions. TSM acknowledges funding from DST/ICPS/QuST/ 2019/Q67 and I-HUB QTF. MN is supported by Xiamen University Malaysia Research Fund (Grant No. XMUMRF/2020-C5/IMAT/0013). YKL is supported by Xiamen University Malaysia Research Fund (Grant no.~XMUMRF/2021-C8/IPHY/0001).
\appendix
\section{Relation Between Spin Coherence and Zeros of Bivariate LY Polynomials} \label{appenA}
The initial state of the probe and system is given in Eq.~\Eqref{eq:initial-state} in main text. We evolve this state to get $\rho(t) = \mathcal{U}(t)\rho(0)\mathcal{U}^{\dagger}(t)$, where $\mathcal{U}(t) = \exp{(-\im tH/\hbar)}$ for the Hamiltonian $H$ given in Eq.~\Eqref{eq:Hint} in main text :
\begin{figure*}
	\centering
	\includegraphics[width = 4.8cm]{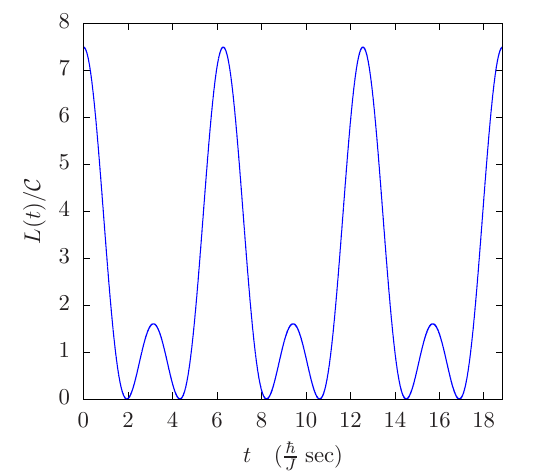} 
	\includegraphics[width = 4cm]{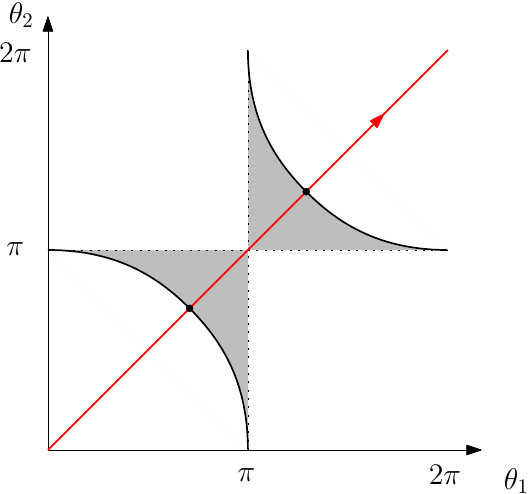}
	\includegraphics[width=4.8cm]{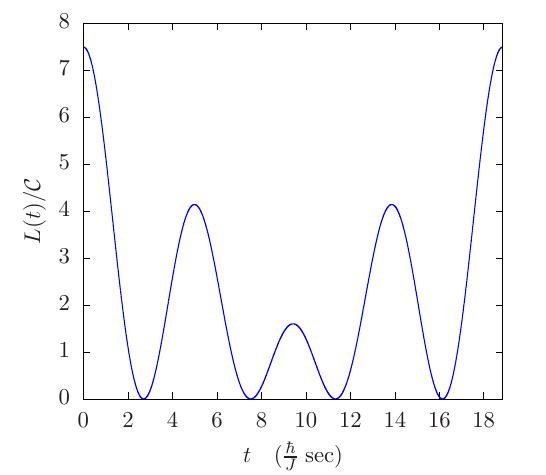} 
	\includegraphics[width=4cm]{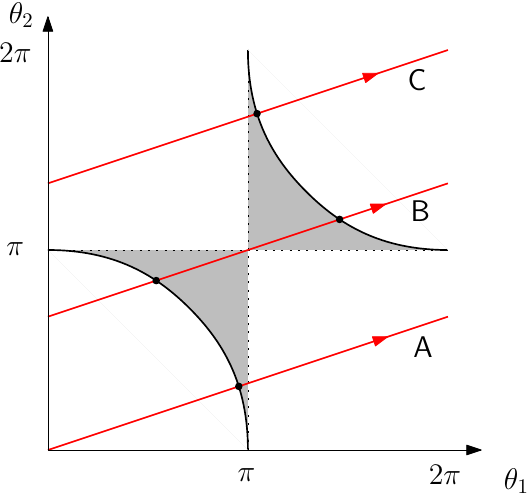} 	
	\hspace{0.5cm} (a) \hspace{3.7cm} (b) \hspace{4.2cm} (c) \hspace {3.8cm} (d) \\  
	$m = 1$  \hspace{8cm} $m = \frac{1}{3}$
	\caption{The sampling of a \emph{coamoeba} for $\lambda_A=J/4$, $\lambda_B=mJ/4$, and $\beta=0.5$, for (a-b) $m=1$, and (c-d) $m=\frac{1}{3}$, plotted for $0\leq t\leq6\pi\frac{\hbar}{J}$. (a) and (c) show the plots of $L(t)$ vs $t$, while (b) and (d) depict the parameterized lines Eq.~\Eqref{eq:coam} in the main text, overlaid with a sketch of the \emph{coamoeba}. The points when $L(t)=0$ correspond to the intersection of the lines with the boundary of the \emph{coamoeba}.}
	\label{fig:rational}
\end{figure*}

\begin{align}
	\rho_P(t) & = \mathrm{tr}_{AB} \left( \mathcal{U}(t) ~ \rho(0) ~ \mathcal{U}^{\dagger}(t) \right)
	\nonumber \\
	& =  \frac{ \langle 0 | \psi_P\rangle \langle \psi_P | 0 \rangle ~ + ~ \langle 1 | \psi_P \rangle \langle \psi_P | 1 \rangle  }{Z(\beta,h_A,h_B)} \nonumber \\
	& + \frac{ \langle 0 | \psi_P \rangle \langle \psi_P | 1 \rangle }{Z(\beta,h_A,h_B)} [ \mathrm{e}^{\beta(h_A+h_B+J)} \mathrm{e}^{-2\im(\lambda_A + \lambda_B)t/\hbar} \nonumber\\
	&+  \mathrm{e}^{\beta(h_A-h_B-J)} \mathrm{e}^{-2\im(\lambda_A - \lambda_B)t/\hbar} \nonumber \\
	&  +\mathrm{e}^{\beta(-h_A+h_B-J)} \mathrm{e}^{-2\im(-\lambda_A + \lambda_B)t/\hbar} \nonumber \\
	&+  e^{\beta(-h_A-h_B+J)} \mathrm{e}^{-2\im(-\lambda_A - \lambda_B)t/\hbar}] + \mathrm{h.c}  \nonumber
\end{align}
From this we get (using the complexified variables $z_1$,~$z_2$ and $\Gamma$ defined in the main text) : 
\begin{align}
	\langle \sigma_x^P (t) \rangle = \frac{ \langle 0 | \psi_P \rangle \langle \psi_P | 1 \rangle }{Z(\beta,h_A,h_B) (\Gamma z_1 z_2)^{\frac{1}{2}}} \nonumber \\
	\left( 1 + \Gamma z_1 + \Gamma z_2 + z_1 z_2 \right) + \mathrm{h.c} \label{eq:sigmax} \\
	\mathrm{and} ~~  \langle \sigma_y^P (t) \rangle = \im\frac{ \langle 0 | \psi_P \rangle \langle \psi_P | 1 \rangle }{Z(\beta,h_A,h_B) (\Gamma z_1 z_2)^{\frac{1}{2}}} \nonumber \\
	\left( 1 + \Gamma z_1 + \Gamma z_2 + z_1 z_2 \right) + \mathrm{h.c} \label{eq:sigmay}. 	
\end{align}	
Therefore, we get the probe coherence $L$ as a function of time as :
\begin{align}
	L(t) = & \frac{\hbar^2}{4} |\langle \sigma_x^P (t) \rangle + \im\langle \sigma_y^P (t) \rangle |^2 \nonumber \\
	& = \frac{\hbar^2 |\langle 1 | \psi_P \rangle \langle \psi_P | 0 \rangle|^2 }{Z^{2}(\beta,h_A,h_B) (\Gamma z_1^{*} z_2^{*})} |1 + \Gamma z_1^* + \Gamma z_2^* + z_1^* z_2^*|^2 \nonumber \\
	& = \mathcal{C} |f(z_1,z_2)|^2,
\end{align}
where, $\mathcal{C} = \frac{\hbar^2 |\langle 1 | \psi_P \rangle \langle \psi_P | 0 \rangle|^2 }{Z^{2}(\beta,h_A,h_B) (\Gamma z_1^{*} z_2^{*})} $, and $f(z_1,z_2) = 1 + \Gamma z_1 + \Gamma z_2 + z_1 z_2$ is a bivariate LY polynomial. Hence the derivation of Eq.~\Eqref{eq:Lt} is complete.
\section{Sampling of Coamoeba: Methodology} \label{AppenB}
 Since, at points where the LY polynomial $|f(z_1,z_2)|^2=0$, the arguments of the variables represent points of the \emph{coamoeba}, Eq.~\Eqref{eq:coam} in the main text are straight lines in the coamoeba plane, with slope $\lambda_B/\lambda_A$, parametrised by $t$. As the time $t$ elapses, a point traverses the coamoeba plane along a straight line at velocity $\brac{4\lambda_A/\hbar,4\lambda_B/\hbar}$. The spin coherence $L(t)$ of Eq.~\Eqref{eq:Lt} in the main text vanishes whenever $(\theta_1(t),\theta_2(t))$ coincides with the corresponding point(s) on the coamoeba.

\begin{figure}
		\centering
		\includegraphics[width=4.2cm]{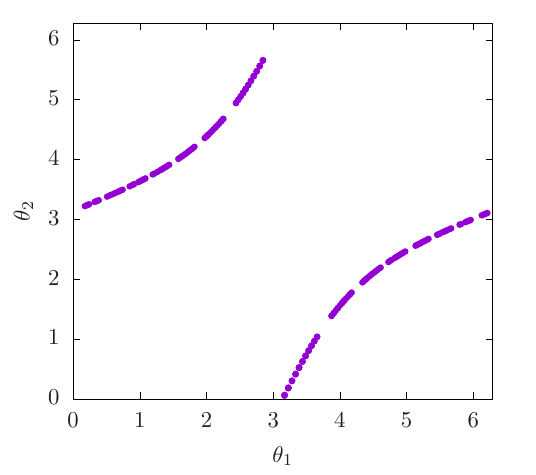}
		\includegraphics[width=4.2cm]{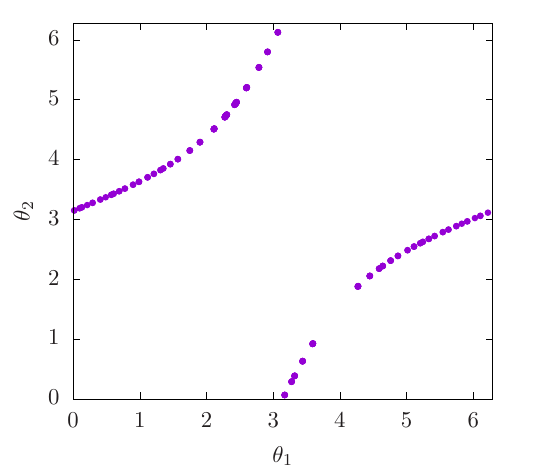} \\
		(a) \hspace{4cm} (b)
		\label{fig_Jnet_msqrt2}
	\caption{Physical sampling of a coamoeba in the anti-ferromagnetic case, $J<0$, where $\lambda_A=-J/4$, $\lambda_B=-mJ/4$, and $\beta=0.5$ for (a) $m = \pi$ and (b) $m = \sqrt{2}$. The points are obtained by recording the times where $L(t)$ vanishes in the interval $0\leq t\leq 500\hbar/|J|$.}
	\label{fig_Jneg}
\end{figure}

As an example, we plot in Fig.~\ref{fig:rational} $L(t)$ vs $t$, along with the parametric lines $\brac{\theta_1(t),\theta_2(t)}$ for the case $\lambda_A=J/4$, $\lambda_B=mJ/4$, $\beta=0.5$, and $h_A=h_B=0$;  for the cases $m=1$ and $m=\frac{1}{3}$. We see that $L(t)$ vanishes whenever the line $\brac{\theta_1(t),\theta_2(t)}$ intersects the  \emph{coamoeba} boundary. For the case $m=1$ (Fig.~\ref{fig:rational}~(b)), the line only intersects the  \emph{coamoeba} boundary twice before it repeats after a period of $2\pi\frac{\hbar}{J}$. For $m=\frac{1}{3}$ (Fig.~\ref{fig:rational}~(d)), the line starts from the origin as segment \textsf{A} where it intersects the  \emph{coamoeba} once, then continues as segment \textsf{B}, intersecting the  \emph{coamoeba} twice, and finally as segment \textsf{C}, where it intersects the  \emph{coamoeba} one more time before it repeats the segments \textsf{A}, \textsf{B}, \textsf{C} after a period of $6\pi\frac{\hbar}{J}$.
\begin{figure*}
	\centering
	\includegraphics[width=4.4cm]{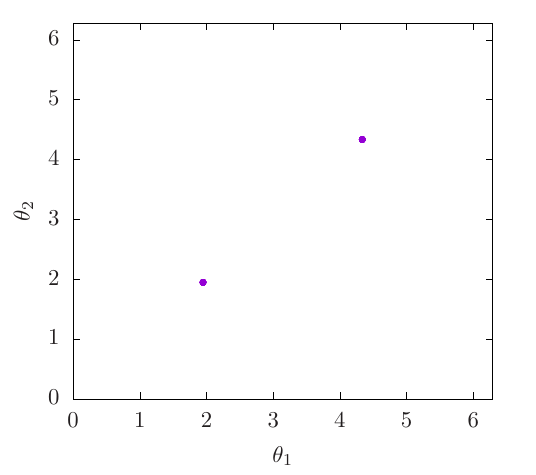}
	\includegraphics[width=4.4cm]{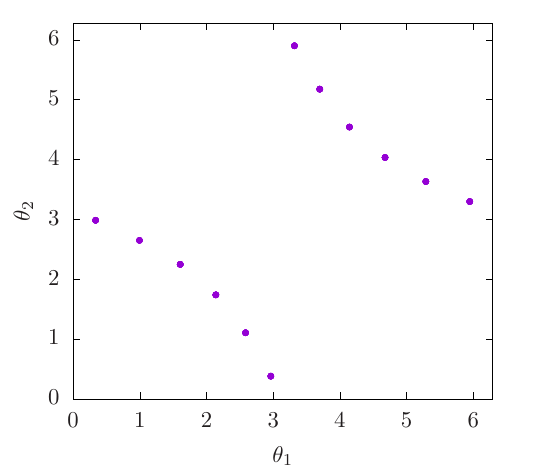}
	\includegraphics[width=4.4cm]{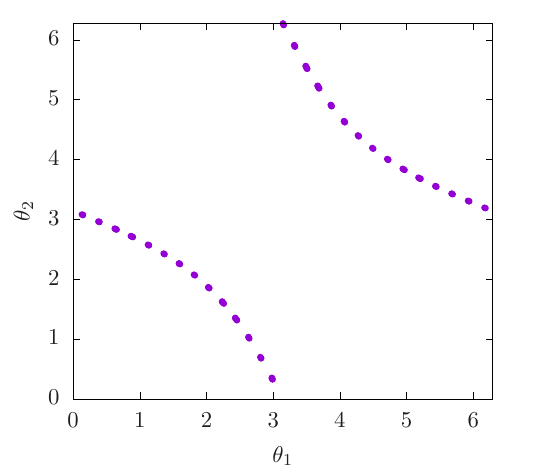}
	\includegraphics[width=4.4cm]{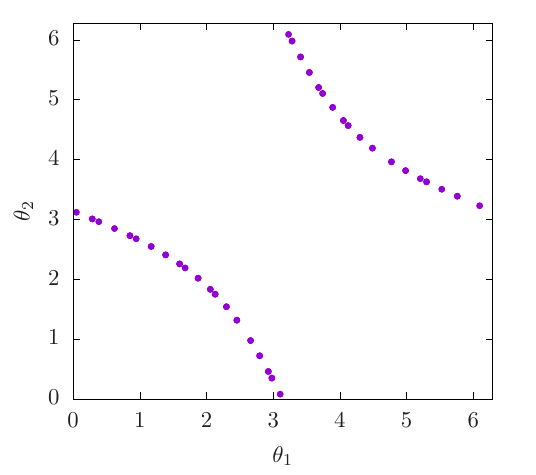}
	\hspace{0.5cm} (a) \hspace{3.7cm} (b) \hspace{4.2cm} (c) \hspace {3.8cm} (d) \\
	$m=1$ \hspace{3.2cm} $m=\frac{7}{5}$ \hspace{4cm} $m=\pi$ \hspace {3cm} $m=\sqrt{2}$
	\label{fig_Jpos_msqrt2}
	\caption{Physical sampling of a coamoeba in the ferromagnetic case, $J>0$, with $\lambda_A=J/4$, $\lambda_B=mJ/4$, and $\beta=0.5$ for various $m$. The points are obtained by recording the times where $L(t)$ vanishes in the interval $0\leq t\leq 100\hbar/J$.}
	\label{fig_Jpos}
\end{figure*}
Thus, an experimental procedure to sample a  \emph{coamoeba} can be done as follows. A two-spin system is prepared in a heat bath of temperature $1/\beta$, and coherence $L(t)$ of a quantum probe coupled to the system is measured. We then record the times when $L(t)$ vanishes. This gives a sequence, say, $t_1,t_2,t_3,\ldots$. By Eq.~\Eqref{eq:coam} in the main text, this will form a collection of points on the torus. As explained above, these mark the intersection points of the line with the coamoeba. As long as $m=\frac{\lambda_B}{\lambda_A}$ is not rational, (which, in an experimental situation, is most likely the case) we can eventually collect enough points to form the shape of the coamoeba. Figure \ref{fig_Jpos} demonstrates this for the ferromagnetic case $J>0$ with $t$ running in the interval $0\leq t\leq 100\hbar/J$. The anti-ferromagnetic case is shown in Fig.~\ref{fig_Jneg}, where $t$ is $0\leq t\leq 500\hbar/|J|$.
In our case, $m$ is irrational and equals to $J_{PB}/J_{PA}$, whose values are listed in Fig.~\ref{fig:molecule}~(b) in the main text. 
\section{Experiments for $h_A = h_B = 0$ } \label{AppendC}
\subsection{Preparation via Probe :} \label{AppendC1}
By setting $h_A=h_B=0$ in Eq.~\eqref{eq:target}, we get the target state 
\begin{equation}
	\rho^\Delta (0) = \cosh(\beta J) I_x^P + \sinh(\beta J) 4I_x^P I_z^A I_z^B. \label{eq:ImTar}
\end{equation}  
Here we have dropped the proportionality factor of $1/Z(\beta,0,0)$. To prepare this, we start with the NMR thermal deviation state $\rho_\mathrm{th}^{\Delta}(0)$ as mentioned in the main text and proceed : 
\begin{align}
I_z^P \xrightarrow{(\pi/2)^P_y} I_x^P \xrightarrow{\tau} \cos(\pi J_{PA} \tau) \cos(\pi J_{PB} \tau) I_x^P + \nonumber\\
\cos(\pi J_{PA} \tau) \sin(\pi J_{PB} \tau) 2 I_y^P I_z^B  \nonumber \\
+\sin(\pi J_{PA} \tau)  \cos(\pi J_{PB} \tau) 2I_y^P I_z^A \nonumber\\
- \sin(\pi J_{PA} \tau)\sin(\pi J_{PB} \tau) 4I_x^P I_z^A I_z^B \nonumber \\ 
\downarrow{(\pi/2)^P_y - \mathrm{Grad.} - (\pi/2)^P_{-y}} \nonumber \\
\cos(\pi J_{PA} \tau)\cos(\pi J_{PB} \tau) I_x^P \nonumber\\
 - \sin(\pi J_{PA} \tau)\sin(\pi J_{PB} \tau) 4I_x^P I_z^A I_z^B \label{eq:Imprep}
\end{align} 
Thus the target state is achieved by applying pulses on the probe alone considering the system to be inaccessible. Here, by $\tau$ we meant free evolution under $H_{\mathrm{eff}}$ (given in the main text) and Grad. represents a pulsed filed gradient pulse along the $z$-direction. In particular, the prepared state of Eq.~\Eqref{eq:Imprep} equals the target state of Eq.~\Eqref{eq:ImTar} (apart from trivial proportionality factors) when 
\begin{align}
	\frac{\cos(\pi(J_{PA}+J_{PB})\tau)}{\cos(\pi(J_{PA}-J_{PB})\tau)} = \expo{2\beta J} \label{rq:ImCond}.
\end{align}
Therefore for any value of $\beta J$, we can find the corresponding $\tau$ satisfying the above equation. Setting that delay time $\tau$ in the pulse sequence, the initial state can be readily prepared. For example, we prepare the state for $\beta J = 1/2$ by setting $\tau = 8.7646484$ ms. In this value of $\tau$, the fidelity between Eq.~\Eqref{eq:ImTar} and Eq.~\Eqref{eq:Imprep} becomes $0.99$. On the other hand, to initiate the state for $\beta J = -1/2$, we set $\tau=9.0136719$ ms in which case the respective fidelity is $\approx 1$.
\subsection{Analysis :} \label{AppendC2}
To sample the \emph{coamoeba} we initiate the system as mentioned above both for ferromagnetic ($\beta J = 1/2$) and anti-ferromagnetic ($\beta J = -1/2$) cases. After the initiation, as mentioned in the main text,  we just need to let it evolve under its NMR internal effective Hamiltonian $H_{\mathrm{eff}}$ (form is given in the main text). By direct computation, we get
\begin{align}
\langle \sigma_x^P (t) \rangle  = \mathrm{tr} [\sigma_x^P \expo{-\im H_{\mathrm{eff}}t/\hbar} (	\cosh(\beta J) I_x^P \nonumber\\
+ \sinh(\beta J) 4I_x^P I_z^A I_z^B) \expo{\im H_{\mathrm{eff}}t/\hbar}] \nonumber \\
\propto \left\{ \expo{\beta J} \cos[\pi(J_{PA} + J_{PB})t] + \expo{-\beta J} \cos[\pi(J_{PA} - J_{PB})t]\right\}, \label{eq:Imreal} \\
\mathrm{and}, ~~  \langle \sigma_y^P (t) \rangle  = \mathrm{tr} [\sigma_y^P \expo{-\im H_{\mathrm{eff}}t/\hbar} (	\cosh(\beta J) I_x^P \nonumber\\
+ \sinh(\beta J) 4I_x^P I_z^A I_z^B) \expo{\im H_{\mathrm{eff}}t/\hbar}] = 0. 
\end{align}

\begin{figure}
	\centering
	\includegraphics[height = 3.8cm, trim={3cm 0cm 3cm 0cm},  clip=true]{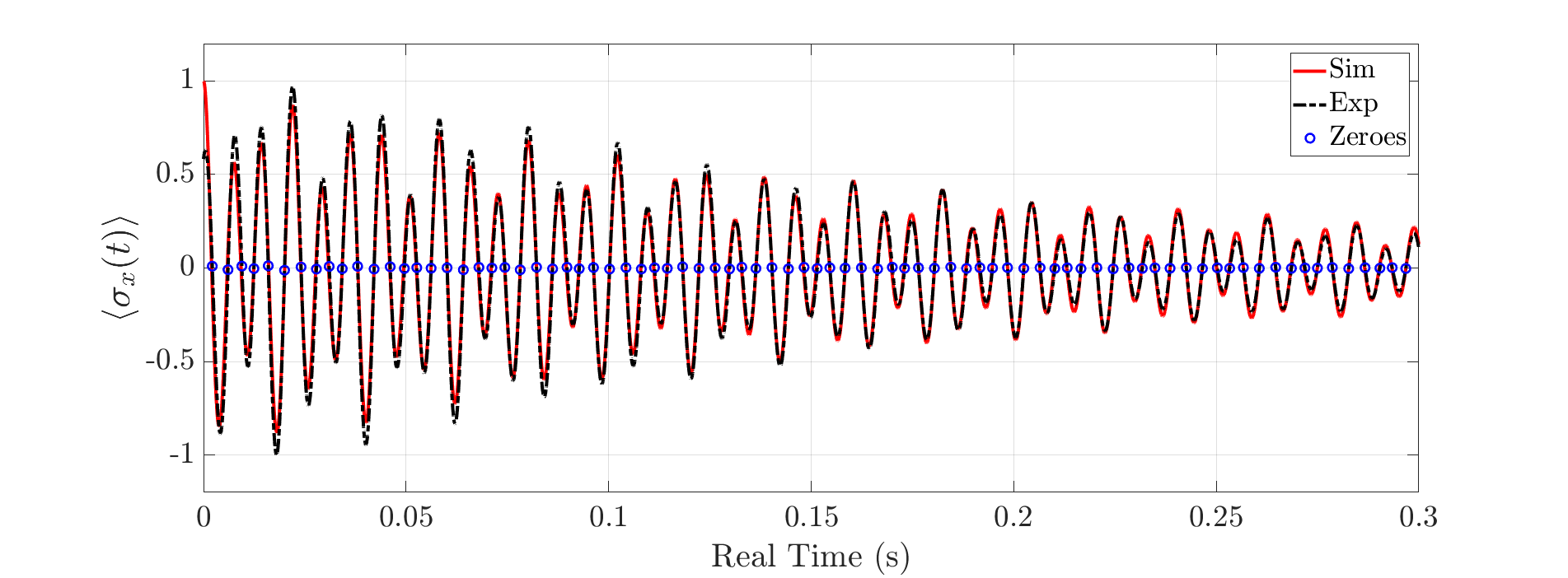} \\
	(a)
	\includegraphics[height = 3.2cm, trim={3cm 0cm 3cm 0cm},  clip=true]{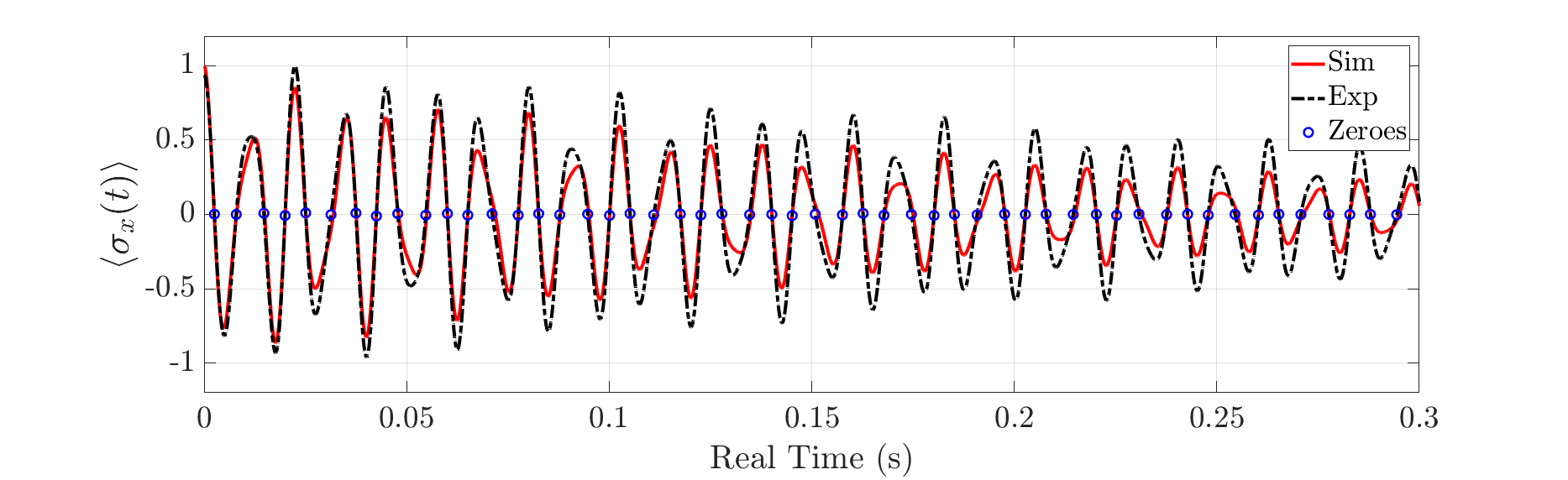} \\
	(b) 
	\caption{(a) Simulated and Experimental FID for $\beta J = 1/2$. (b) Same for $\beta J = -1/2$. Zeros of FID are marked with circles} \label{ImFID}
\end{figure}

As $\langle \sigma_y^P (t) \rangle$ is zero throughout,  we just need to find the the time points where $\langle \sigma_x^P (t) \rangle$ vanishes. In Fig.~\ref{ImFID}, we plot the real part of direct NMR FID on top of the predicted FID by analytical expression of Eq~\Eqref{eq:Imreal} and observe they match really well. (To correct the initial phase error due to electronic switching time etc, we have performed a zeroth order phase correction on the data). Experimentally observed time points, where the coherence vanishes, are noted and mapped to the coamoeba torus via Eq.~\Eqref{eq:coam} of the main text.
\section{Experiments for $h_A,h_B~\neq~0$} \label{AppendD}
\subsection{Preparation via Probe :} \label{AppendD1}
To initialize the system at any arbitrary non-zero value of $h_A$ and $h_B$, full state mentioned in Eq.~\Eqref{eq:target} becomes our target. Following the sequences of pulses given in Fig~\ref{fig:CompExp}~(a) in the main text, we achieve this preparation. Here we prove how the target is achieved by the pulse sequence of Fig.~\ref{fig:CompExp}~(a), starting from thermal equilibrium :
\begin{align}
	I_z^P \xrightarrow{\theta^1_y} \cos(\theta^1)I_z^P + \sin(\theta^1) I_x^P 
	\xrightarrow{\theta^2_{-x}} \cos(\theta^1)\cos(\theta^2) I_z^P \nonumber\\
	+  \cos(\theta^1)\sin(\theta^2) I_y^P +  \sin(\theta^1) I_x^P \nonumber \\
	\downarrow{\tau} \nonumber \\
	\cos(\theta^1)\cos(\theta^2) I_z^P + 	\cos(\theta^1)\sin(\theta^2) 
	 [\cos(\pi J_{PA} \tau)\nonumber\\
	 \cos(\pi J_{PB} \tau) I_y^P  - \cos(\pi J_{PA} \tau)\sin(\pi J_{PB} \tau) 2I_x^PI_z^B \nonumber \\
	 - \sin(\pi J_{PA} \tau)\cos(\pi J_{PB} \tau) 2I_x^PI_z^A - \sin(\pi J_{PA} \tau) \nonumber\\ 
	 \sin(\pi J_{PB} \tau) 4I_y^PI_z^AI_z^B] +
	  \sin(\theta^1)[\cos(\pi J_{PA} \tau) \nonumber \\
	  \cos(\pi J_{PB} \tau) I_x^P 
	  + \cos(\pi J_{PA} \tau)\sin(\pi J_{PB} \tau) 2I_y^PI_z^B \nonumber \\
	  + \sin(\pi J_{PA} \tau)\cos(\pi J_{PB} \tau) 2I_y^PI_z^A - \sin(\pi J_{PA} \tau)  \nonumber\\ 
	  \sin(\pi J_{PB} \tau) 4I_x^PI_z^AI_z^B] \nonumber \\
	 \downarrow{(\pi/2)^P_y - \mathrm{Grad.} - (\pi/2)^P_{-y}} 
	 \nonumber \\
	 [\sin(\theta^1)\cos(\pi J_{PA} \tau)\cos(\pi J_{PB} \tau)] I_x^P \nonumber \\
	  +  [-\cos(\theta^1)\sin(\theta^2)\sin(\pi J_{PA} \tau)\cos(\pi J_{PB} \tau)] 2I_x^PI_z^A \nonumber \\
	 [-\cos(\theta^1)\sin(\theta^2)\cos(\pi J_{PA} \tau)\sin(\pi J_{PB} \tau)] 2I_x^PI_z^B \nonumber \\
	  +[-\sin(\theta^1)  \sin(\pi J_{PA} \tau)\sin(\pi J_{PB} \tau)] 4I_x^PI_z^AI_z^B   \label{eq:compPrep}.
\end{align}
Comparing this prepared state with the target state of Eq.~\Eqref{eq:target}, we note that to prepare for a given value of $\{\beta h_A ,\beta h_B ,\beta J\}$, we need to find corresponding values of free parameters $\{\theta^1,\theta^2,\tau\}$ from the bellow four equations :
\begin{subequations}
	\begin{align}
	C_a =   [\sin(\theta^1)\cos(\pi J_{PA} \tau)\cos(\pi J_{PB} \tau)]  \label{opa} \\
	C_b= [-\cos(\theta^1)\sin(\theta^2)\sin(\pi J_{PA} \tau)\cos(\pi J_{PB} \tau)]  \label{opb} \\ 
	C_c=  [-\cos(\theta^1)\sin(\theta^2)\cos(\pi J_{PA} \tau)\sin(\pi J_{PB} \tau)] \label{opc} \\ 
	C_d= [-\sin(\theta^1)  \sin(\pi J_{PA} \tau)\sin(\pi J_{PB} \tau)]  \label{opd}
	\end{align}
\end{subequations}
Therefore it boils down to a three parameter estimation problem, given a set of four equations. In particular, if we denote right hand sides of Eq.~\Eqref{opa}\Eqref{opb}\Eqref{opc}\Eqref{opd} as $\{c_a,c_b,c_c,c_d\}$, respectively, then we can define the optimization problem as :
\begin{align}
	\mathrm{Optimize~for} ~ \{\theta^1,\theta^2,\tau\} ~~\mathrm{such~that,} \nonumber \\
	f = |\vec{C}-\vec{c}| = \sum_{i=\{a,b,c,d\}} |C_i - c_i| ~\mathrm{is~minimized.}  
\end{align}        
Any available optimization algorithm can be employed to minimize $f$. For example, we have used genetic algorithm. Of-course, as the system gets larger the optimization problem will become difficult unless the system posses some symmetry. However, our method allows us to reduce this optimization problem into two optimization problems with lesser complexity. To explain how that can be done we compute $\langle \sigma_x^P (t) \rangle$ and  $\langle \sigma_y^P (t) \rangle$ as the state of Eq.~\Eqref{eq:target} evolves under $\mathcal{U}(t) = \exp(-\im H_{\mathrm{eff}}t/\hbar)$:
\begin{subequations}
	\begin{align}
		&\langle \sigma_x^P (t) \rangle = \mathrm{tr}[\mathcal{U}\rho_\Delta(0)\mathcal{U}^{\dagger} \sigma_x] \nonumber \\
		&= C_a \cos(\pi J_{PA}t)\cos(\pi J_{PB}t) - C_d \sin(\pi J_{PA}t)\sin(\pi J_{PB}t) \label{lopf}\\
		&\langle \sigma_y^P (t) \rangle = \mathrm{tr}[\mathcal{U}\rho_\Delta(0)\mathcal{U}^{\dagger} \sigma_y] \nonumber \\
		&= C_b \sin(\pi J_{PA}t)\cos(\pi J_{PB}t) + C_c \cos(\pi J_{PA}t)\sin(\pi J_{PB}t). \label{lops}
	\end{align}
\end{subequations}
From Eq.~\Eqref{lopf} and Eq.~\Eqref{lops}, we note that only $c_a$ and $c_d$ contributes in the measurement of $\langle \sigma_x^P (t) \rangle$ while $c_b$ and $c_c$ term contributes in $\langle \sigma_y^P (t) \rangle$ measurement. This mutual exclusivity can be exploited to prepare Eq.~\Eqref{eq:target} via optimizing for first and last term alone without caring for the other two terms and measuring $\langle \sigma_x^P (t) \rangle$. Then, by the same logic, we optimize only for the second and third term of  Eq.~\Eqref{eq:target} and measure  $\langle \sigma_y^P (t) \rangle$.

As mentioned in the main text, we perform our experiments at two non-zero values of $h_A$ and $h_B$, namely  $\beta J = 1/2, \beta h_A = \beta h_B = 0.1$ and $\beta J = 1/2, \beta h_A = - \beta h_B = 0.1$.  For each of these cases,  we perform two sets of experiments by extracting $\langle \sigma_x^P(t) \rangle$ and $\langle \sigma_y^P(t) \rangle$ separately to demonstrate the before-mentioned optimization-splitting.  
Using the optimized values, pulse sequence of Fig.~\ref{fig:CompExp}~(b) in the main text is employed for state initialization. After initialization is achieved, $\langle \sigma_x^P(t) \rangle$ and $\langle \sigma_y^P (t) \rangle$ are recorded as the NMR register evolves freely under $H_{\mathrm{eff}}$ (form is given in the main text). After zeroth order phase correction on the experimental data, interpolated signals are plotted in Fig.~\ref{fig:CompExp}~(g-h) in the main text. Their zero points (marked by circles) are then mapped to the 2-torus by Eq.~\Eqref{eq:coam} in the main text. 

\nocite{*}
\bibliography{bibliography}

\begin{thebibliography}{31}%
\makeatletter
\providecommand \@ifxundefined [1]{%
 \@ifx{#1\undefined}
}%
\providecommand \@ifnum [1]{%
 \ifnum #1\expandafter \@firstoftwo
 \else \expandafter \@secondoftwo
 \fi
}%
\providecommand \@ifx [1]{%
 \ifx #1\expandafter \@firstoftwo
 \else \expandafter \@secondoftwo
 \fi
}%
\providecommand \natexlab [1]{#1}%
\providecommand \enquote  [1]{``#1''}%
\providecommand \bibnamefont  [1]{#1}%
\providecommand \bibfnamefont [1]{#1}%
\providecommand \citenamefont [1]{#1}%
\providecommand \href@noop [0]{\@secondoftwo}%
\providecommand \href [0]{\begingroup \@sanitize@url \@href}%
\providecommand \@href[1]{\@@startlink{#1}\@@href}%
\providecommand \@@href[1]{\endgroup#1\@@endlink}%
\providecommand \@sanitize@url [0]{\catcode `\\12\catcode `\$12\catcode
  `\&12\catcode `\#12\catcode `\^12\catcode `\_12\catcode `\%12\relax}%
\providecommand \@@startlink[1]{}%
\providecommand \@@endlink[0]{}%
\providecommand \url  [0]{\begingroup\@sanitize@url \@url }%
\providecommand \@url [1]{\endgroup\@href {#1}{\urlprefix }}%
\providecommand \urlprefix  [0]{URL }%
\providecommand \Eprint [0]{\href }%
\providecommand \doibase [0]{http://dx.doi.org/}%
\providecommand \selectlanguage [0]{\@gobble}%
\providecommand \bibinfo  [0]{\@secondoftwo}%
\providecommand \bibfield  [0]{\@secondoftwo}%
\providecommand \translation [1]{[#1]}%
\providecommand \BibitemOpen [0]{}%
\providecommand \bibitemStop [0]{}%
\providecommand \bibitemNoStop [0]{.\EOS\space}%
\providecommand \EOS [0]{\spacefactor3000\relax}%
\providecommand \BibitemShut  [1]{\csname bibitem#1\endcsname}%
\let\auto@bib@innerbib\@empty
\bibitem [{\citenamefont {Yang}\ and\ \citenamefont {Lee}(1952)}]{Yang:1952be}%
  \BibitemOpen
  \bibfield  {author} {\bibinfo {author} {\bibfnamefont {C.-N.}\ \bibnamefont
  {Yang}}\ and\ \bibinfo {author} {\bibfnamefont {T.~D.}\ \bibnamefont {Lee}},\
  }\href {\doibase 10.1103/PhysRev.87.404} {\bibfield  {journal} {\bibinfo
  {journal} {Phys. Rev.}\ }\textbf {\bibinfo {volume} {87}},\ \bibinfo {pages}
  {404} (\bibinfo {year} {1952})}\BibitemShut {NoStop}%
\bibitem [{\citenamefont {Lee}\ and\ \citenamefont {Yang}(1952)}]{Lee:1952ig}%
  \BibitemOpen
  \bibfield  {author} {\bibinfo {author} {\bibfnamefont {T.~D.}\ \bibnamefont
  {Lee}}\ and\ \bibinfo {author} {\bibfnamefont {C.-N.}\ \bibnamefont {Yang}},\
  }\href {\doibase 10.1103/PhysRev.87.410} {\bibfield  {journal} {\bibinfo
  {journal} {Phys. Rev.}\ }\textbf {\bibinfo {volume} {87}},\ \bibinfo {pages}
  {410} (\bibinfo {year} {1952})}\BibitemShut {NoStop}%
\bibitem [{\citenamefont {Francis}\ \emph {et~al.}(2021)\citenamefont
  {Francis}, \citenamefont {Zhu}, \citenamefont {Alderete}, \citenamefont
  {Johri}, \citenamefont {Xiao}, \citenamefont {Freericks}, \citenamefont
  {Monroe}, \citenamefont {Linke},\ and\ \citenamefont
  {Kemper}}]{doi:10.1126/sciadv.abf2447}%
  \BibitemOpen
  \bibfield  {author} {\bibinfo {author} {\bibfnamefont {A.}~\bibnamefont
  {Francis}}, \bibinfo {author} {\bibfnamefont {D.}~\bibnamefont {Zhu}},
  \bibinfo {author} {\bibfnamefont {C.~H.}\ \bibnamefont {Alderete}}, \bibinfo
  {author} {\bibfnamefont {S.}~\bibnamefont {Johri}}, \bibinfo {author}
  {\bibfnamefont {X.}~\bibnamefont {Xiao}}, \bibinfo {author} {\bibfnamefont
  {J.~K.}\ \bibnamefont {Freericks}}, \bibinfo {author} {\bibfnamefont
  {C.}~\bibnamefont {Monroe}}, \bibinfo {author} {\bibfnamefont {N.~M.}\
  \bibnamefont {Linke}}, \ and\ \bibinfo {author} {\bibfnamefont {A.~F.}\
  \bibnamefont {Kemper}},\ }\href {\doibase 10.1126/sciadv.abf2447} {\bibfield
  {journal} {\bibinfo  {journal} {Science Advances}\ }\textbf {\bibinfo
  {volume} {7}},\ \bibinfo {pages} {eabf2447} (\bibinfo {year} {2021})},\
  \Eprint
  {http://arxiv.org/abs/https://www.science.org/doi/pdf/10.1126/sciadv.abf2447}
  {https://www.science.org/doi/pdf/10.1126/sciadv.abf2447} \BibitemShut
  {NoStop}%
\bibitem [{\citenamefont {Deger}\ and\ \citenamefont
  {Flindt}(2019)}]{PhysRevResearch.1.023004}%
  \BibitemOpen
  \bibfield  {author} {\bibinfo {author} {\bibfnamefont {A.}~\bibnamefont
  {Deger}}\ and\ \bibinfo {author} {\bibfnamefont {C.}~\bibnamefont {Flindt}},\
  }\href {\doibase 10.1103/PhysRevResearch.1.023004} {\bibfield  {journal}
  {\bibinfo  {journal} {Phys. Rev. Res.}\ }\textbf {\bibinfo {volume} {1}},\
  \bibinfo {pages} {023004} (\bibinfo {year} {2019})}\BibitemShut {NoStop}%
\bibitem [{\citenamefont {Brandner}\ \emph {et~al.}(2017)\citenamefont
  {Brandner}, \citenamefont {Maisi}, \citenamefont {Pekola}, \citenamefont
  {Garrahan},\ and\ \citenamefont {Flindt}}]{PhysRevLett.118.180601}%
  \BibitemOpen
  \bibfield  {author} {\bibinfo {author} {\bibfnamefont {K.}~\bibnamefont
  {Brandner}}, \bibinfo {author} {\bibfnamefont {V.~F.}\ \bibnamefont {Maisi}},
  \bibinfo {author} {\bibfnamefont {J.~P.}\ \bibnamefont {Pekola}}, \bibinfo
  {author} {\bibfnamefont {J.~P.}\ \bibnamefont {Garrahan}}, \ and\ \bibinfo
  {author} {\bibfnamefont {C.}~\bibnamefont {Flindt}},\ }\href {\doibase
  10.1103/PhysRevLett.118.180601} {\bibfield  {journal} {\bibinfo  {journal}
  {Phys. Rev. Lett.}\ }\textbf {\bibinfo {volume} {118}},\ \bibinfo {pages}
  {180601} (\bibinfo {year} {2017})}\BibitemShut {NoStop}%
\bibitem [{\citenamefont {Deger}\ \emph {et~al.}(2018)\citenamefont {Deger},
  \citenamefont {Brandner},\ and\ \citenamefont {Flindt}}]{PhysRevE.97.012115}%
  \BibitemOpen
  \bibfield  {author} {\bibinfo {author} {\bibfnamefont {A.}~\bibnamefont
  {Deger}}, \bibinfo {author} {\bibfnamefont {K.}~\bibnamefont {Brandner}}, \
  and\ \bibinfo {author} {\bibfnamefont {C.}~\bibnamefont {Flindt}},\ }\href
  {\doibase 10.1103/PhysRevE.97.012115} {\bibfield  {journal} {\bibinfo
  {journal} {Phys. Rev. E}\ }\textbf {\bibinfo {volume} {97}},\ \bibinfo
  {pages} {012115} (\bibinfo {year} {2018})}\BibitemShut {NoStop}%
\bibitem [{\citenamefont {Bastid}\ \emph {et~al.}(2005)\citenamefont {Bastid},
  \citenamefont {Andronic}, \citenamefont {Barret}, \citenamefont {Basrak},
  \citenamefont {Benabderrahmane}, \citenamefont {\ifmmode~\check{C}\else
  \v{C}\fi{}aplar}, \citenamefont {Cordier}, \citenamefont {Crochet},
  \citenamefont {Dupieux}, \citenamefont {D\ifmmode~\check{z}\else
  \v{z}\fi{}elalija}, \citenamefont {Fodor}, \citenamefont {Ga\ifmmode
  \check{s}\else \v{s}\fi{}pari\ifmmode~\acute{c}\else \'{c}\fi{}},
  \citenamefont {Gobbi}, \citenamefont {Grishkin}, \citenamefont {Hartmann},
  \citenamefont {Herrmann}, \citenamefont {Hildenbrand}, \citenamefont {Hong},
  \citenamefont {Kecskemeti}, \citenamefont {Kim}, \citenamefont {Kirejczyk},
  \citenamefont {Koczon}, \citenamefont {Korolija}, \citenamefont {Kotte},
  \citenamefont {Kress}, \citenamefont {Lebedev}, \citenamefont {Leifels},
  \citenamefont {Lopez}, \citenamefont {Mangiarotti}, \citenamefont {Manko},
  \citenamefont {Merschmeyer}, \citenamefont {Moisa}, \citenamefont {Neubert},
  \citenamefont {Pelte}, \citenamefont {Petrovici}, \citenamefont {Rami},
  \citenamefont {Reisdorf}, \citenamefont {Schuettauf}, \citenamefont {Seres},
  \citenamefont {Sikora}, \citenamefont {Sim}, \citenamefont {Simion},
  \citenamefont {Siwek-Wilczy\ifmmode~\acute{n}\else \'{n}\fi{}ska},
  \citenamefont {Smolarkiewicz}, \citenamefont {Smolyankin}, \citenamefont
  {Soliwoda}, \citenamefont {Stockmeier}, \citenamefont {Stoicea},
  \citenamefont {Tyminski}, \citenamefont {Wi\ifmmode~\acute{s}\else
  \'{s}\fi{}niewski}, \citenamefont {Wohlfarth}, \citenamefont {Xiao},
  \citenamefont {Yushmanov}, \citenamefont {Zhilin}, \citenamefont
  {Ollitrault},\ and\ \citenamefont {Borghini}}]{PhysRevC.72.011901}%
  \BibitemOpen
  \bibfield  {author} {\bibinfo {author} {\bibfnamefont {N.}~\bibnamefont
  {Bastid}}, \bibinfo {author} {\bibfnamefont {A.}~\bibnamefont {Andronic}},
  \bibinfo {author} {\bibfnamefont {V.}~\bibnamefont {Barret}}, \bibinfo
  {author} {\bibfnamefont {Z.}~\bibnamefont {Basrak}}, \bibinfo {author}
  {\bibfnamefont {M.~L.}\ \bibnamefont {Benabderrahmane}}, \bibinfo {author}
  {\bibfnamefont {R.}~\bibnamefont {\ifmmode~\check{C}\else \v{C}\fi{}aplar}},
  \bibinfo {author} {\bibfnamefont {E.}~\bibnamefont {Cordier}}, \bibinfo
  {author} {\bibfnamefont {P.}~\bibnamefont {Crochet}}, \bibinfo {author}
  {\bibfnamefont {P.}~\bibnamefont {Dupieux}}, \bibinfo {author} {\bibfnamefont
  {M.}~\bibnamefont {D\ifmmode~\check{z}\else \v{z}\fi{}elalija}}, \bibinfo
  {author} {\bibfnamefont {Z.}~\bibnamefont {Fodor}}, \bibinfo {author}
  {\bibfnamefont {I.}~\bibnamefont {Ga\ifmmode \check{s}\else
  \v{s}\fi{}pari\ifmmode~\acute{c}\else \'{c}\fi{}}}, \bibinfo {author}
  {\bibfnamefont {A.}~\bibnamefont {Gobbi}}, \bibinfo {author} {\bibfnamefont
  {Y.}~\bibnamefont {Grishkin}}, \bibinfo {author} {\bibfnamefont {O.~N.}\
  \bibnamefont {Hartmann}}, \bibinfo {author} {\bibfnamefont {N.}~\bibnamefont
  {Herrmann}}, \bibinfo {author} {\bibfnamefont {K.~D.}\ \bibnamefont
  {Hildenbrand}}, \bibinfo {author} {\bibfnamefont {B.}~\bibnamefont {Hong}},
  \bibinfo {author} {\bibfnamefont {J.}~\bibnamefont {Kecskemeti}}, \bibinfo
  {author} {\bibfnamefont {Y.~J.}\ \bibnamefont {Kim}}, \bibinfo {author}
  {\bibfnamefont {M.}~\bibnamefont {Kirejczyk}}, \bibinfo {author}
  {\bibfnamefont {P.}~\bibnamefont {Koczon}}, \bibinfo {author} {\bibfnamefont
  {M.}~\bibnamefont {Korolija}}, \bibinfo {author} {\bibfnamefont
  {R.}~\bibnamefont {Kotte}}, \bibinfo {author} {\bibfnamefont
  {T.}~\bibnamefont {Kress}}, \bibinfo {author} {\bibfnamefont
  {A.}~\bibnamefont {Lebedev}}, \bibinfo {author} {\bibfnamefont
  {Y.}~\bibnamefont {Leifels}}, \bibinfo {author} {\bibfnamefont
  {X.}~\bibnamefont {Lopez}}, \bibinfo {author} {\bibfnamefont
  {A.}~\bibnamefont {Mangiarotti}}, \bibinfo {author} {\bibfnamefont
  {V.}~\bibnamefont {Manko}}, \bibinfo {author} {\bibfnamefont
  {M.}~\bibnamefont {Merschmeyer}}, \bibinfo {author} {\bibfnamefont
  {D.}~\bibnamefont {Moisa}}, \bibinfo {author} {\bibfnamefont
  {W.}~\bibnamefont {Neubert}}, \bibinfo {author} {\bibfnamefont
  {D.}~\bibnamefont {Pelte}}, \bibinfo {author} {\bibfnamefont
  {M.}~\bibnamefont {Petrovici}}, \bibinfo {author} {\bibfnamefont
  {F.}~\bibnamefont {Rami}}, \bibinfo {author} {\bibfnamefont {W.}~\bibnamefont
  {Reisdorf}}, \bibinfo {author} {\bibfnamefont {A.}~\bibnamefont
  {Schuettauf}}, \bibinfo {author} {\bibfnamefont {Z.}~\bibnamefont {Seres}},
  \bibinfo {author} {\bibfnamefont {B.}~\bibnamefont {Sikora}}, \bibinfo
  {author} {\bibfnamefont {K.~S.}\ \bibnamefont {Sim}}, \bibinfo {author}
  {\bibfnamefont {V.}~\bibnamefont {Simion}}, \bibinfo {author} {\bibfnamefont
  {K.}~\bibnamefont {Siwek-Wilczy\ifmmode~\acute{n}\else \'{n}\fi{}ska}},
  \bibinfo {author} {\bibfnamefont {M.~M.}\ \bibnamefont {Smolarkiewicz}},
  \bibinfo {author} {\bibfnamefont {V.}~\bibnamefont {Smolyankin}}, \bibinfo
  {author} {\bibfnamefont {I.~J.}\ \bibnamefont {Soliwoda}}, \bibinfo {author}
  {\bibfnamefont {M.~R.}\ \bibnamefont {Stockmeier}}, \bibinfo {author}
  {\bibfnamefont {G.}~\bibnamefont {Stoicea}}, \bibinfo {author} {\bibfnamefont
  {Z.}~\bibnamefont {Tyminski}}, \bibinfo {author} {\bibfnamefont
  {K.}~\bibnamefont {Wi\ifmmode~\acute{s}\else \'{s}\fi{}niewski}}, \bibinfo
  {author} {\bibfnamefont {D.}~\bibnamefont {Wohlfarth}}, \bibinfo {author}
  {\bibfnamefont {Z.}~\bibnamefont {Xiao}}, \bibinfo {author} {\bibfnamefont
  {I.}~\bibnamefont {Yushmanov}}, \bibinfo {author} {\bibfnamefont
  {A.}~\bibnamefont {Zhilin}}, \bibinfo {author} {\bibfnamefont {J.-Y.}\
  \bibnamefont {Ollitrault}}, \ and\ \bibinfo {author} {\bibfnamefont
  {N.}~\bibnamefont {Borghini}} (\bibinfo {collaboration} {FOPI
  Collaboration}),\ }\href {\doibase 10.1103/PhysRevC.72.011901} {\bibfield
  {journal} {\bibinfo  {journal} {Phys. Rev. C}\ }\textbf {\bibinfo {volume}
  {72}},\ \bibinfo {pages} {011901} (\bibinfo {year} {2005})}\BibitemShut
  {NoStop}%
\bibitem [{\citenamefont {Arndt}\ \emph {et~al.}(2001)\citenamefont {Arndt},
  \citenamefont {Dahmen},\ and\ \citenamefont
  {Hinrichsen}}]{arndt2001directed}%
  \BibitemOpen
  \bibfield  {author} {\bibinfo {author} {\bibfnamefont {P.}~\bibnamefont
  {Arndt}}, \bibinfo {author} {\bibfnamefont {S.}~\bibnamefont {Dahmen}}, \
  and\ \bibinfo {author} {\bibfnamefont {H.}~\bibnamefont {Hinrichsen}},\
  }\href@noop {} {\bibfield  {journal} {\bibinfo  {journal} {Physica A:
  Statistical Mechanics and its Applications}\ }\textbf {\bibinfo {volume}
  {295}},\ \bibinfo {pages} {128} (\bibinfo {year} {2001})}\BibitemShut
  {NoStop}%
\bibitem [{\citenamefont {Krasnytska}\ \emph {et~al.}(2015)\citenamefont
  {Krasnytska}, \citenamefont {Berche}, \citenamefont {Holovatch},\ and\
  \citenamefont {Kenna}}]{krasnytska2015violation}%
  \BibitemOpen
  \bibfield  {author} {\bibinfo {author} {\bibfnamefont {M.}~\bibnamefont
  {Krasnytska}}, \bibinfo {author} {\bibfnamefont {B.}~\bibnamefont {Berche}},
  \bibinfo {author} {\bibfnamefont {Y.}~\bibnamefont {Holovatch}}, \ and\
  \bibinfo {author} {\bibfnamefont {R.}~\bibnamefont {Kenna}},\ }\href@noop {}
  {\bibfield  {journal} {\bibinfo  {journal} {Europhysics Letters}\ }\textbf
  {\bibinfo {volume} {111}},\ \bibinfo {pages} {60009} (\bibinfo {year}
  {2015})}\BibitemShut {NoStop}%
\bibitem [{\citenamefont {Krasnytska}\ \emph {et~al.}(2016)\citenamefont
  {Krasnytska}, \citenamefont {Berche}, \citenamefont {Holovatch},\ and\
  \citenamefont {Kenna}}]{krasnytska2016partition}%
  \BibitemOpen
  \bibfield  {author} {\bibinfo {author} {\bibfnamefont {M.}~\bibnamefont
  {Krasnytska}}, \bibinfo {author} {\bibfnamefont {B.}~\bibnamefont {Berche}},
  \bibinfo {author} {\bibfnamefont {Y.}~\bibnamefont {Holovatch}}, \ and\
  \bibinfo {author} {\bibfnamefont {R.}~\bibnamefont {Kenna}},\ }\href@noop {}
  {\bibfield  {journal} {\bibinfo  {journal} {Journal of Physics A:
  Mathematical and Theoretical}\ }\textbf {\bibinfo {volume} {49}},\ \bibinfo
  {pages} {135001} (\bibinfo {year} {2016})}\BibitemShut {NoStop}%
\bibitem [{\citenamefont {Lee}(2013{\natexlab{a}})}]{PhysRevLett.110.248101}%
  \BibitemOpen
  \bibfield  {author} {\bibinfo {author} {\bibfnamefont {J.}~\bibnamefont
  {Lee}},\ }\href {\doibase 10.1103/PhysRevLett.110.248101} {\bibfield
  {journal} {\bibinfo  {journal} {Phys. Rev. Lett.}\ }\textbf {\bibinfo
  {volume} {110}},\ \bibinfo {pages} {248101} (\bibinfo {year}
  {2013}{\natexlab{a}})}\BibitemShut {NoStop}%
\bibitem [{\citenamefont {Lee}(2013{\natexlab{b}})}]{PhysRevE.88.022710}%
  \BibitemOpen
  \bibfield  {author} {\bibinfo {author} {\bibfnamefont {J.}~\bibnamefont
  {Lee}},\ }\href {\doibase 10.1103/PhysRevE.88.022710} {\bibfield  {journal}
  {\bibinfo  {journal} {Phys. Rev. E}\ }\textbf {\bibinfo {volume} {88}},\
  \bibinfo {pages} {022710} (\bibinfo {year} {2013}{\natexlab{b}})}\BibitemShut
  {NoStop}%
\bibitem [{\citenamefont {Heyl}\ \emph {et~al.}(2013)\citenamefont {Heyl},
  \citenamefont {Polkovnikov},\ and\ \citenamefont
  {Kehrein}}]{PhysRevLett.110.135704}%
  \BibitemOpen
  \bibfield  {author} {\bibinfo {author} {\bibfnamefont {M.}~\bibnamefont
  {Heyl}}, \bibinfo {author} {\bibfnamefont {A.}~\bibnamefont {Polkovnikov}}, \
  and\ \bibinfo {author} {\bibfnamefont {S.}~\bibnamefont {Kehrein}},\ }\href
  {\doibase 10.1103/PhysRevLett.110.135704} {\bibfield  {journal} {\bibinfo
  {journal} {Phys. Rev. Lett.}\ }\textbf {\bibinfo {volume} {110}},\ \bibinfo
  {pages} {135704} (\bibinfo {year} {2013})}\BibitemShut {NoStop}%
\bibitem [{\citenamefont {Flindt}\ and\ \citenamefont
  {Garrahan}(2013)}]{PhysRevLett.110.050601}%
  \BibitemOpen
  \bibfield  {author} {\bibinfo {author} {\bibfnamefont {C.}~\bibnamefont
  {Flindt}}\ and\ \bibinfo {author} {\bibfnamefont {J.~P.}\ \bibnamefont
  {Garrahan}},\ }\href {\doibase 10.1103/PhysRevLett.110.050601} {\bibfield
  {journal} {\bibinfo  {journal} {Phys. Rev. Lett.}\ }\textbf {\bibinfo
  {volume} {110}},\ \bibinfo {pages} {050601} (\bibinfo {year}
  {2013})}\BibitemShut {NoStop}%
\bibitem [{\citenamefont {Dorner}\ \emph {et~al.}(2013)\citenamefont {Dorner},
  \citenamefont {Clark}, \citenamefont {Heaney}, \citenamefont {Fazio},
  \citenamefont {Goold},\ and\ \citenamefont
  {Vedral}}]{PhysRevLett.110.230601}%
  \BibitemOpen
  \bibfield  {author} {\bibinfo {author} {\bibfnamefont {R.}~\bibnamefont
  {Dorner}}, \bibinfo {author} {\bibfnamefont {S.~R.}\ \bibnamefont {Clark}},
  \bibinfo {author} {\bibfnamefont {L.}~\bibnamefont {Heaney}}, \bibinfo
  {author} {\bibfnamefont {R.}~\bibnamefont {Fazio}}, \bibinfo {author}
  {\bibfnamefont {J.}~\bibnamefont {Goold}}, \ and\ \bibinfo {author}
  {\bibfnamefont {V.}~\bibnamefont {Vedral}},\ }\href {\doibase
  10.1103/PhysRevLett.110.230601} {\bibfield  {journal} {\bibinfo  {journal}
  {Phys. Rev. Lett.}\ }\textbf {\bibinfo {volume} {110}},\ \bibinfo {pages}
  {230601} (\bibinfo {year} {2013})}\BibitemShut {NoStop}%
\bibitem [{\citenamefont {Mazzola}\ \emph {et~al.}(2013)\citenamefont
  {Mazzola}, \citenamefont {De~Chiara},\ and\ \citenamefont
  {Paternostro}}]{PhysRevLett.110.230602}%
  \BibitemOpen
  \bibfield  {author} {\bibinfo {author} {\bibfnamefont {L.}~\bibnamefont
  {Mazzola}}, \bibinfo {author} {\bibfnamefont {G.}~\bibnamefont {De~Chiara}},
  \ and\ \bibinfo {author} {\bibfnamefont {M.}~\bibnamefont {Paternostro}},\
  }\href {\doibase 10.1103/PhysRevLett.110.230602} {\bibfield  {journal}
  {\bibinfo  {journal} {Phys. Rev. Lett.}\ }\textbf {\bibinfo {volume} {110}},\
  \bibinfo {pages} {230602} (\bibinfo {year} {2013})}\BibitemShut {NoStop}%
\bibitem [{\citenamefont {Wei}\ \emph {et~al.}(2014)\citenamefont {Wei},
  \citenamefont {Chen}, \citenamefont {Po},\ and\ \citenamefont
  {Liu}}]{wei2014phase}%
  \BibitemOpen
  \bibfield  {author} {\bibinfo {author} {\bibfnamefont {B.-B.}\ \bibnamefont
  {Wei}}, \bibinfo {author} {\bibfnamefont {S.-W.}\ \bibnamefont {Chen}},
  \bibinfo {author} {\bibfnamefont {H.-C.}\ \bibnamefont {Po}}, \ and\ \bibinfo
  {author} {\bibfnamefont {R.-B.}\ \bibnamefont {Liu}},\ }\href@noop {}
  {\bibfield  {journal} {\bibinfo  {journal} {Scientific reports}\ }\textbf
  {\bibinfo {volume} {4}},\ \bibinfo {pages} {5202} (\bibinfo {year}
  {2014})}\BibitemShut {NoStop}%
\bibitem [{\citenamefont {Wei}\ and\ \citenamefont
  {Liu}(2012)}]{PhysRevLett.109.185701}%
  \BibitemOpen
  \bibfield  {author} {\bibinfo {author} {\bibfnamefont {B.-B.}\ \bibnamefont
  {Wei}}\ and\ \bibinfo {author} {\bibfnamefont {R.-B.}\ \bibnamefont {Liu}},\
  }\href {\doibase 10.1103/PhysRevLett.109.185701} {\bibfield  {journal}
  {\bibinfo  {journal} {Phys. Rev. Lett.}\ }\textbf {\bibinfo {volume} {109}},\
  \bibinfo {pages} {185701} (\bibinfo {year} {2012})}\BibitemShut {NoStop}%
\bibitem [{\citenamefont {Peng}\ \emph {et~al.}(2015)\citenamefont {Peng},
  \citenamefont {Zhou}, \citenamefont {Wei}, \citenamefont {Cui}, \citenamefont
  {Du},\ and\ \citenamefont {Liu}}]{PhysRevLett.114.010601}%
  \BibitemOpen
  \bibfield  {author} {\bibinfo {author} {\bibfnamefont {X.}~\bibnamefont
  {Peng}}, \bibinfo {author} {\bibfnamefont {H.}~\bibnamefont {Zhou}}, \bibinfo
  {author} {\bibfnamefont {B.-B.}\ \bibnamefont {Wei}}, \bibinfo {author}
  {\bibfnamefont {J.}~\bibnamefont {Cui}}, \bibinfo {author} {\bibfnamefont
  {J.}~\bibnamefont {Du}}, \ and\ \bibinfo {author} {\bibfnamefont {R.-B.}\
  \bibnamefont {Liu}},\ }\href {\doibase 10.1103/PhysRevLett.114.010601}
  {\bibfield  {journal} {\bibinfo  {journal} {Phys. Rev. Lett.}\ }\textbf
  {\bibinfo {volume} {114}},\ \bibinfo {pages} {010601} (\bibinfo {year}
  {2015})}\BibitemShut {NoStop}%
\bibitem [{\citenamefont {Koll{\'a}r}(2001)}]{kollar2001simplest}%
  \BibitemOpen
  \bibfield  {author} {\bibinfo {author} {\bibfnamefont {J.}~\bibnamefont
  {Koll{\'a}r}},\ }\href@noop {} {\bibfield  {journal} {\bibinfo  {journal}
  {Bulletin of the American Mathematical Society}\ }\textbf {\bibinfo {volume}
  {38}},\ \bibinfo {pages} {409} (\bibinfo {year} {2001})}\BibitemShut
  {NoStop}%
\bibitem [{\citenamefont {{I.~M.~Gelfand, M.~M.~Kapranov, and
  A.~V.~Zelevinsky}}(1994)}]{GKZ-94}%
  \BibitemOpen
  \bibfield  {author} {\bibinfo {author} {\bibnamefont {{I.~M.~Gelfand,
  M.~M.~Kapranov, and A.~V.~Zelevinsky}}},\ }\href@noop {} {\emph {\bibinfo
  {title} {{Discriminants, Resultants, and Multidimensional Determinants}}}}\
  (\bibinfo  {publisher} {{Birkh\"{a}user}},\ \bibinfo {address} {{Boston}},\
  \bibinfo {year} {(1994)})\BibitemShut {NoStop}%
\bibitem [{\citenamefont {Viro}(2002)}]{Viro:2002}%
  \BibitemOpen
  \bibfield  {author} {\bibinfo {author} {\bibfnamefont {O.}~\bibnamefont
  {Viro}},\ }\href@noop {} {\bibfield  {journal} {\bibinfo  {journal} {Notices
  of the AMS}\ }\textbf {\bibinfo {volume} {49}},\ \bibinfo {pages} {916 }
  (\bibinfo {year} {2002})}\BibitemShut {NoStop}%
\bibitem [{\citenamefont {Feng}\ \emph {et~al.}(2008)\citenamefont {Feng},
  \citenamefont {He}, \citenamefont {Kennaway},\ and\ \citenamefont
  {Vafa}}]{Feng:2005gw}%
  \BibitemOpen
  \bibfield  {author} {\bibinfo {author} {\bibfnamefont {B.}~\bibnamefont
  {Feng}}, \bibinfo {author} {\bibfnamefont {Y.-H.}\ \bibnamefont {He}},
  \bibinfo {author} {\bibfnamefont {K.~D.}\ \bibnamefont {Kennaway}}, \ and\
  \bibinfo {author} {\bibfnamefont {C.}~\bibnamefont {Vafa}},\ }\href {\doibase
  10.4310/ATMP.2008.v12.n3.a2} {\bibfield  {journal} {\bibinfo  {journal} {Adv.
  Theor. Math. Phys.}\ }\textbf {\bibinfo {volume} {12}},\ \bibinfo {pages}
  {489} (\bibinfo {year} {2008})},\ \Eprint
  {http://arxiv.org/abs/hep-th/0511287} {arXiv:hep-th/0511287} \BibitemShut
  {NoStop}%
\bibitem [{\citenamefont {Nisse}\ and\ \citenamefont
  {Sottile}(2013{\natexlab{a}})}]{Nisse:2013a}%
  \BibitemOpen
  \bibfield  {author} {\bibinfo {author} {\bibfnamefont {M.}~\bibnamefont
  {Nisse}}\ and\ \bibinfo {author} {\bibfnamefont {F.}~\bibnamefont
  {Sottile}},\ }\href {\doibase 10.2140/ant.2013.7.339} {\bibfield  {journal}
  {\bibinfo  {journal} {Algebra \& Number Theory}\ }\textbf {\bibinfo {volume}
  {7(2)}},\ \bibinfo {pages} {339} (\bibinfo {year} {2013}{\natexlab{a}})},\
  \Eprint {http://arxiv.org/abs/1106.0096} {arXiv:1106.0096 [math]}
  \BibitemShut {NoStop}%
\bibitem [{\citenamefont {Nisse}\ and\ \citenamefont
  {Sottile}(2013{\natexlab{b}})}]{Nisse:2013b}%
  \BibitemOpen
  \bibfield  {author} {\bibinfo {author} {\bibfnamefont {M.}~\bibnamefont
  {Nisse}}\ and\ \bibinfo {author} {\bibfnamefont {F.}~\bibnamefont
  {Sottile}},\ }\href {\doibase 10.1090/conm/605/12112} {\bibfield  {journal}
  {\bibinfo  {journal} {Contemporary Mathematics}\ }\textbf {\bibinfo {volume}
  {605}},\ \bibinfo {pages} {73} (\bibinfo {year} {2013}{\natexlab{b}})},\
  \Eprint {http://arxiv.org/abs/1110.1033} {arXiv:1110.1033 [math]}
  \BibitemShut {NoStop}%
\bibitem [{\citenamefont {Nisse}\ and\ \citenamefont {Passare}()}]{Nisse:2017}%
  \BibitemOpen
  \bibfield  {author} {\bibinfo {author} {\bibfnamefont {M.}~\bibnamefont
  {Nisse}}\ and\ \bibinfo {author} {\bibfnamefont {M.}~\bibnamefont
  {Passare}},\ }\enquote {\bibinfo {title} {{Amoebas and Coamoebas of Linear
  Spaces}},}\ \Eprint {http://arxiv.org/abs/1205.2808} {arXiv:1205.2808 [math]}
  \BibitemShut {NoStop}%
\bibitem [{\citenamefont {Sadykov}\ and\ \citenamefont {Zhukov}()}]{sadyT}%
  \BibitemOpen
  \bibfield  {author} {\bibinfo {author} {\bibfnamefont {T.}~\bibnamefont
  {Sadykov}}\ and\ \bibinfo {author} {\bibfnamefont {T.}~\bibnamefont
  {Zhukov}},\ }\href {http://amoebas.ru/index.html} {\enquote {\bibinfo {title}
  {{Amoebas~[dot]~ru}},}\ }\bibinfo {note} {Accessed on 6'th October,
  2023}\BibitemShut {NoStop}%
\bibitem [{\citenamefont {Nielsen}\ and\ \citenamefont
  {Chuang}(2010)}]{nielsen2010quantum}%
  \BibitemOpen
  \bibfield  {author} {\bibinfo {author} {\bibfnamefont {M.~A.}\ \bibnamefont
  {Nielsen}}\ and\ \bibinfo {author} {\bibfnamefont {I.~L.}\ \bibnamefont
  {Chuang}},\ }\href@noop {} {\emph {\bibinfo {title} {Quantum computation and
  quantum information}}}\ (\bibinfo  {publisher} {Cambridge university press},\
  \bibinfo {year} {2010})\BibitemShut {NoStop}%
\bibitem [{\citenamefont {{J.~Cavanagh, W.~J.~Fairbrother, A.~G.~Palmer III,
  and N.~J.~Skelton}}(1996)}]{cavanagh1996protein}%
  \BibitemOpen
  \bibfield  {author} {\bibinfo {author} {\bibnamefont {{J.~Cavanagh,
  W.~J.~Fairbrother, A.~G.~Palmer III, and N.~J.~Skelton}}},\ }\href@noop {}
  {\emph {\bibinfo {title} {{Protein NMR Spectroscopy: Principles and
  Practice}}}}\ (\bibinfo  {publisher} {{Academic Press}},\ \bibinfo {address}
  {{Cambridge, Mass.}},\ \bibinfo {year} {(1996)})\BibitemShut {NoStop}%
\bibitem [{\citenamefont {{M.~H.~Levitt}}(2013)}]{levitt2013spin}%
  \BibitemOpen
  \bibfield  {author} {\bibinfo {author} {\bibnamefont {{M.~H.~Levitt}}},\
  }\href@noop {} {\emph {\bibinfo {title} {{Spin Dynamics: Basics of Nuclear
  Magnetic Resonance}}}}\ (\bibinfo  {publisher} {{John Wiley \& Sons}},\
  \bibinfo {address} {{New York}},\ \bibinfo {year} {(2013)})\BibitemShut
  {NoStop}%
\bibitem [{\citenamefont {Fukushima}(2018)}]{fukushima2018experimental}%
  \BibitemOpen
  \bibfield  {author} {\bibinfo {author} {\bibfnamefont {E.}~\bibnamefont
  {Fukushima}},\ }\href@noop {} {\emph {\bibinfo {title} {Experimental pulse
  NMR: a nuts and bolts approach}}}\ (\bibinfo  {publisher} {CRC Press},\
  \bibinfo {year} {2018})\BibitemShut {NoStop}%
\end{thebibliography}%
	
\end{document}